\documentclass[10pt,a4paper]{article} % 10pt is ignored!
\pdfoutput=1
\usepackage{jcappub}

\allowdisplaybreaks

\usepackage{ifthen}
\usepackage{graphicx}% Include figure files
\usepackage{dcolumn}% Align table columns on decimal point
\usepackage{bm}% bold math
\usepackage{color}
\usepackage{amsmath}
\usepackage{amssymb}

\newcommand{\Mpc}{\ \text{Mpc}}
\newcommand{\hMpc}{\ h^{-1}\text{Mpc}}

\newcommand{\ihMpc}{\ h\text{Mpc}^{-1}}

\newcommand{\hMs}{\ h^{-1} M_\odot}
\newcommand{\eh}[1]{\exp{\left[#1\right]}}
\newcommand{\tim}[1]{\times 10^{#1}}%times 10 

\newcommand{\derd}{\,\mathrm{d}} % upright d for derivatives and integrals
\newcommand{\ddir}{\delta^\text{(D)}}

\newcommand{\be}{\begin{equation}}
\newcommand{\ee}{\end{equation}}
\newcommand{\la}{\left\langle}
\newcommand{\ra}{\right\rangle}
\newcommand{\derivd}{\text{d}}
\renewcommand{\vec}{\bm}
\newcommand{\dqc}{\frac{\derivd^3q}{(2\pi)^3}}

\newcommand{\llg}{\text{l}}
\newcommand{\ii}{\text{i}}

\newcommand{\omnot}{\Omega_\text{m,0}}
\newcommand{\deltal}{\delta_\text{l}}

\date{\today}

\title{Linear response to long wavelength fluctuations using curvature simulations}
\author[a]{Tobias Baldauf,}
\emailAdd{baldauf@ias.edu}
\author[b]{Uro\v{s} Seljak,}
\emailAdd{useljak@berkeley.edu}
\author[c,d]{Leonardo Senatore,}
\emailAdd{senatore@stanford.edu}
\author[a]{Matias Zaldarriaga}
\emailAdd{matiasz@ias.edu}

\affiliation[a]{School of Natural Sciences, Institute for Advanced Study, Princeton, NJ}
\affiliation[b]{Physics Department, Astronomy Department and Lawrence Berkeley National Laboratory, University of California, Berkeley, CA, USA}
\affiliation[c]{Stanford Institute for Theoretical Physics, Stanford University, Stanford, CA, USA}
\affiliation[d]{Kavli Institute for Particle Astrophysics and Cosmology, Menlo Park, CA, USA}

\abstract{
We study the local response to long wavelength fluctuations in cosmological $N$-body simulations, focusing on the matter and halo power spectra, halo abundance and non-linear transformations of the density field. The long wavelength mode is implemented using an effective curved cosmology and a mapping of time and distances. The method provides an alternative, most probably more precise, way to measure the isotropic halo biases. Limiting ourselves to the linear case, we find generally good agreement between the biases obtained from the curvature method and the traditional power spectrum method at the level of a few percent. 
We also study the response of halo counts to changes in the variance of the field and find that the slope of the relation between the responses to density and variance differs from the na\"\i ve derivation assuming a universal mass function by $18\%$. This has implications for measurements of the amplitude of local non-Gaussianity using scale dependent bias. We also analyze the halo power spectrum and halo-dark matter cross-spectrum response to long wavelength fluctuations and derive second order halo bias from it, as well as the super-sample variance
contribution to the galaxy power spectrum covariance matrix.
}
\begin{document}
\maketitle

%===============================================================================
%===============================================================================
\section{Introduction}
One of the most important goals of Large Scale Structure cosmology is to learn about the composition of the Universe and the physical processes that lead to the initial conditions and that are driving its evolution.
Different models for inflation, for dark energy or gravity are usually quantified in terms of the linear power spectrum, though recent analytical efforts have begun to explore the cosmological information contained in the quasi linear dynamics.
Besides gravitational lensing, most observables in the late time Universe are not
directly related to the matter distribution. Yet, they contain a wealth of
cosmological information once the microphysics that cannot be reliably described has been
marginalized over.
In bias models one absorbs the non-perturbative effects that lead for
instance to halo or galaxy formation into a small number of response functions
which then allow the mapping between the tracer clustering properties and the
underlying long wavelength fluctuations. 
Unless one is able to predict the bias coefficients  with some non-perturbative technique, such as numerical simulations, bias in most cases is treated as mere nuisance. However, primordial non-Gaussianities introduce a functional form for bias that can not be produced by a standard cosmological evolution, clearly providing an interesting probe of the dynamics during inflation~(as first proposed by \cite{Dalal:2008hh}).

The purpose of this paper is to discuss galaxy bias by using its interpretation as the response of observables to long wavelength
fluctuations in a different setting. As we have shown in \cite{Baldauf:2011ew}, the effect of
long wavelength modes on local dynamics can be accounted for by an effective
curved local Universe. If the long mode has a larger wavelength than the system
under consideration, the long mode is effectively constant over the relevant
scales. This allows us to treat the long mode as quasi linear, restricting ourselves to isotropic configurations. For simplicity, in this paper we will limit ourselves to treating the long mode at linear level, the generalization to second, third, and higher orders being straightforward \cite{Wagner:2015se,Wagner:2015th}. The short modes, instead, are treated completely non-linear by solving $N$-body simulations.

Another biased tracer is the Ly-$\alpha$ forest, where 
the absorption features are imprinted into the
emission lines of quasars by intervening hydrogen clouds, which is a good
probe of matter fluctuations between $2<z<4$. Here one can also derive the bias as a response to the 
long wavelength fluctuation~\cite{McDonald:2003fx,Seljak:2012ja}.

This paper is a collection of results obtained over the past few years. Several groups have used the observations and formalism developed in~\cite{Baldauf:2011ew} to measure some aspects of the effect of long modes on short modes~\cite{Sirko:2005in,Li:2014ss,Wagner:2015se,Wagner:2015th}. Here in this paper we focus on the original application that was envisioned in~\cite{Baldauf:2011ew}, which is the exploration of statistics related to biased tracers, which have not been discussed so far. 

The benefit of this technique can be understood by comparing it with the standard method which uses the cross correlation between dark matter and tracers at long wavelengths in large simulation boxes of realistic cosmologies. In a realistic cosmology, large scale overdensities are very small and thus shot noise corrections make measurements on the largest scales difficult. If one pushes to smaller scales where the overdensities are larger, one has to worry about perturbative corrections that change the shape of the spectra. In our technique we are considering infinitely long wavelengths for the overdensities but modestly large amplitudes and thus we greatly minimize both shot noise and perturbative corrections.\footnote{These same two observations also suggest that we can measure the bias coefficients by performing small-volume simulations where the fundamental mode is enhanced, and then measuring the distribution of the tracers wavenumber of the enhanced mode. Again, this allows for a measurement of the bias coefficients with no cosmic variance, and small volumes. Contrary to the approach we develop here, this technique also allows one to measure the anisotropic biases.}

This paper is structured as follows: in Sec.~\ref{sec:exp} and~\ref{sec:howtorun} we review and further discuss the mapping between the global and the local Universes and describe the simulations In Sec.~\ref{sec:power} we discuss the matter density perturbations in the local frame, before we describe halo density bias in Sec.~\ref{sec:bias}. Finally, in Sec.~\ref{sec:nltrans} we consider the Ly-alpha flux as an example of non-linear transformations which can be tested using the curvature method before we conclude in Sec.~\ref{sec:concl}.
%===============================================================================
%===============================================================================
\section{Expansion History}\label{sec:exp}
\begin{figure}[h]
\centering
\includegraphics[width=0.49\textwidth]{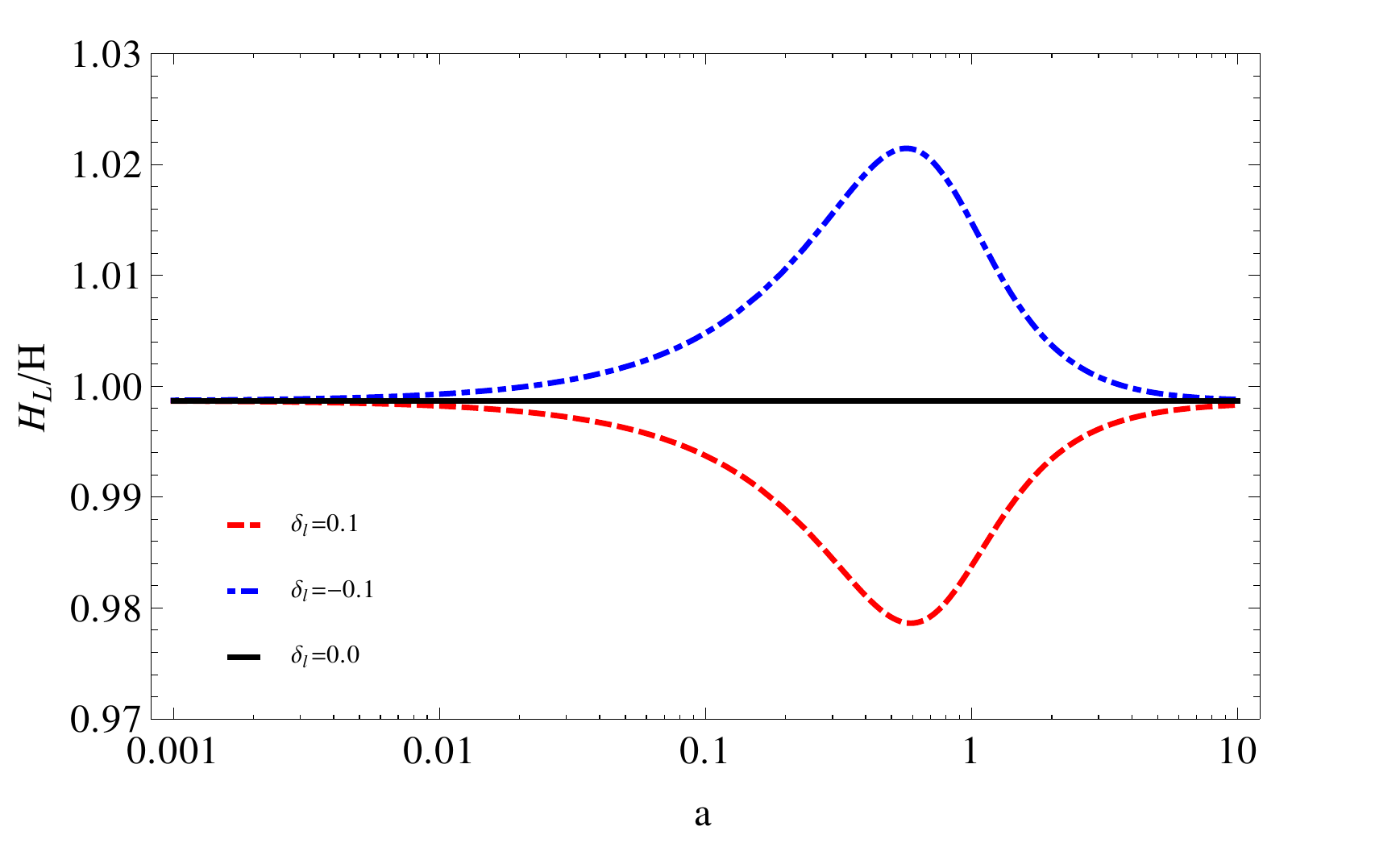}
\includegraphics[width=0.49\textwidth]{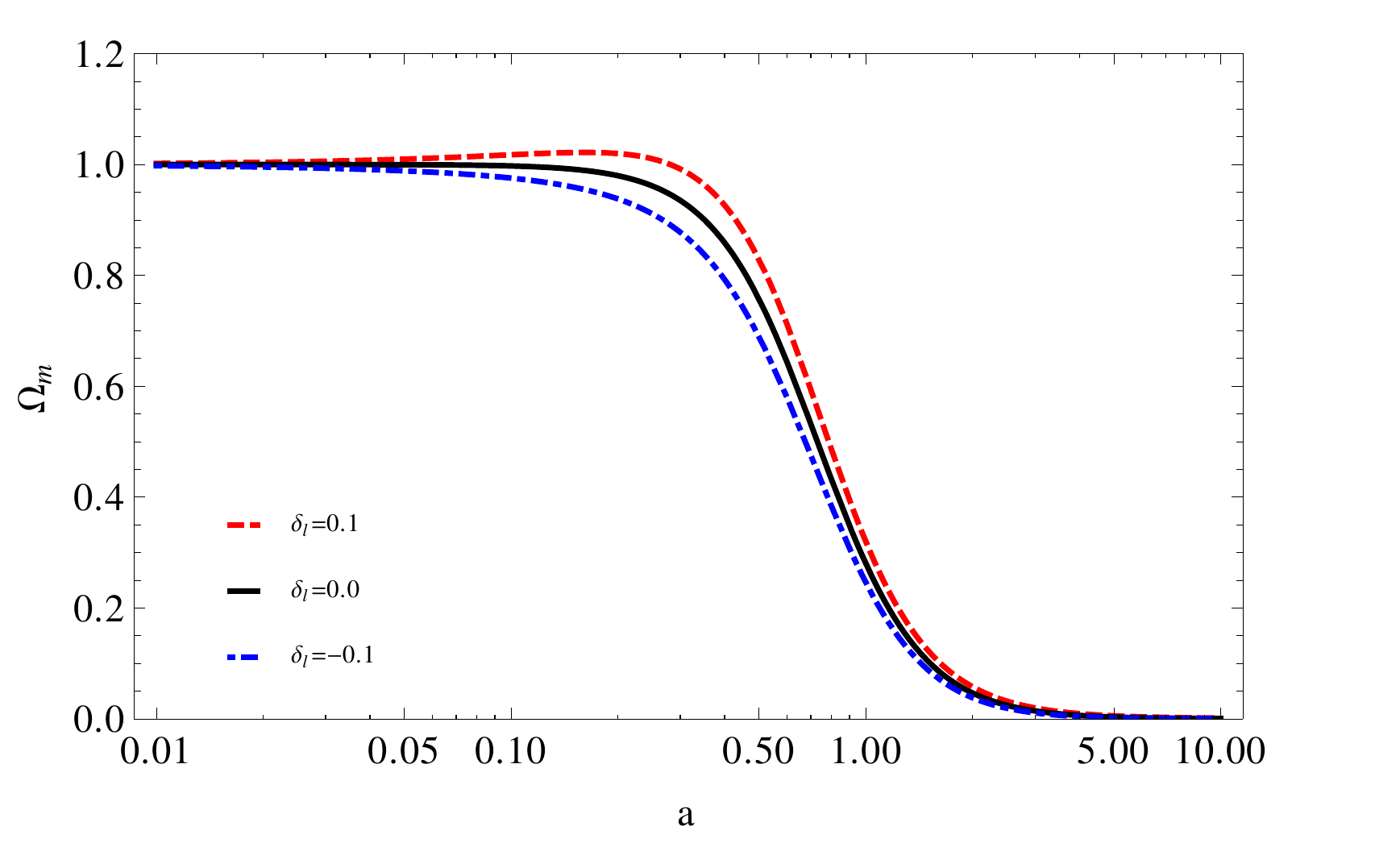}\\
\includegraphics[width=0.49\textwidth]{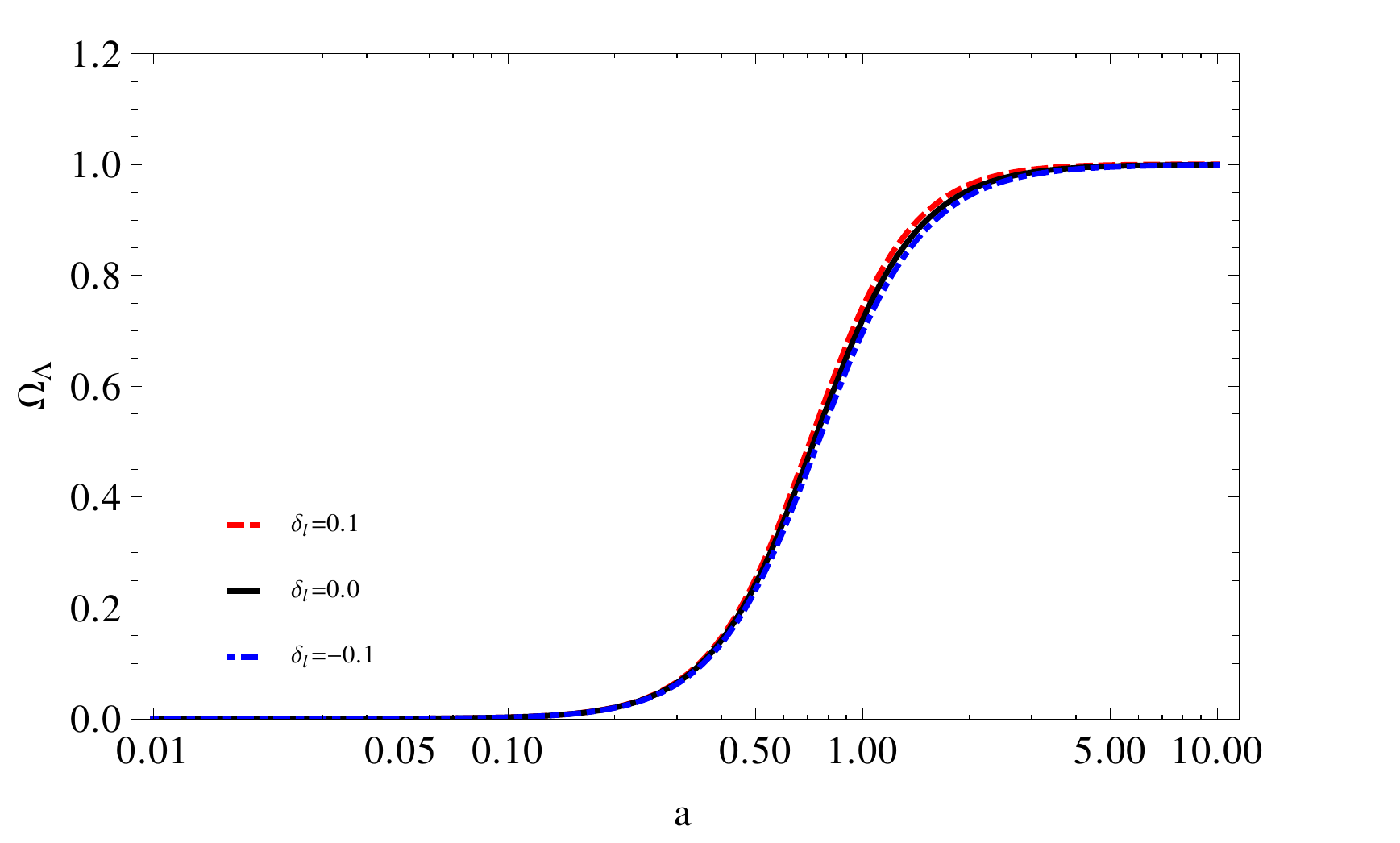}
\includegraphics[width=0.49\textwidth]{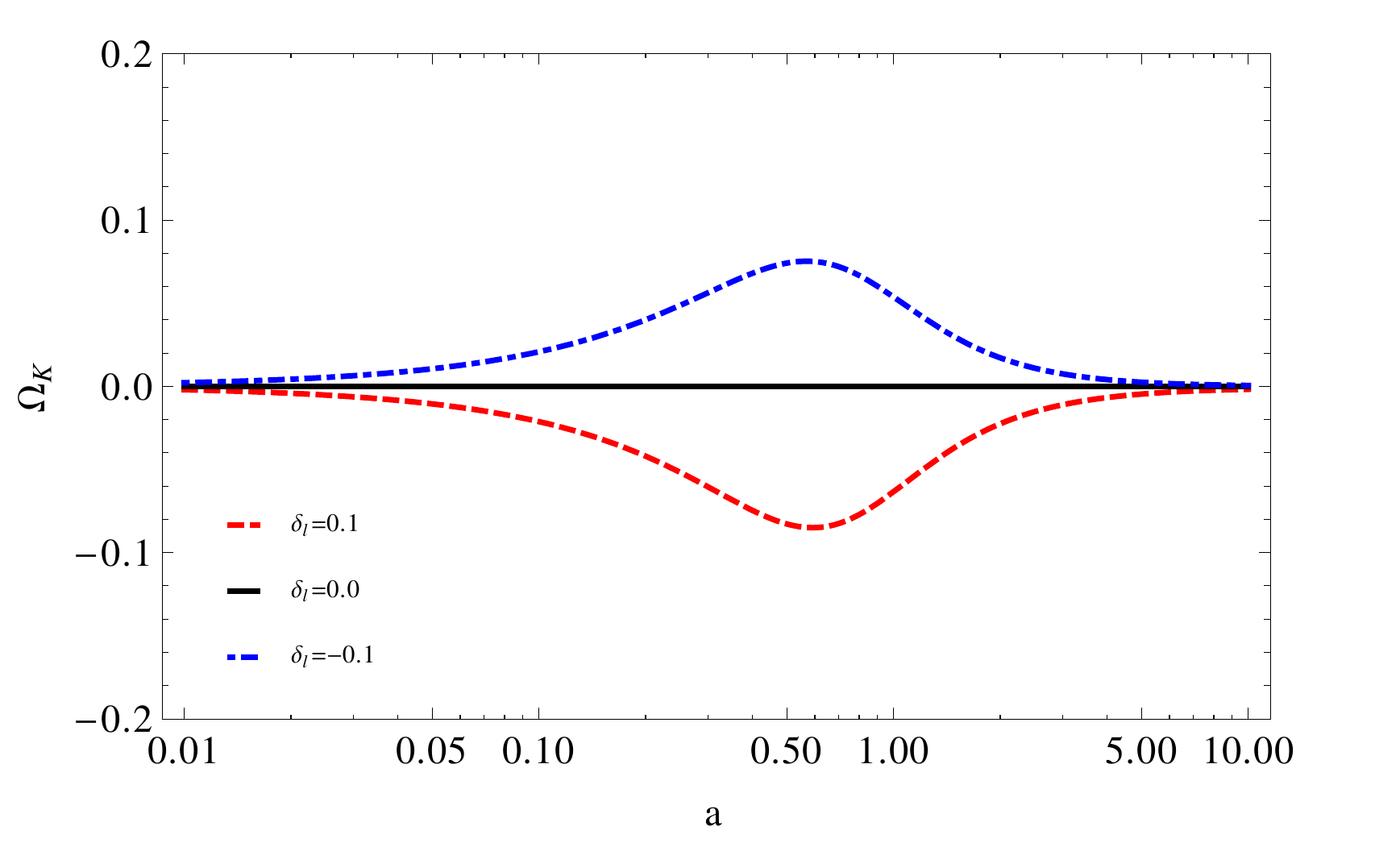}
\caption{Effective expansion history. Red dashed over dense and ble dash-dotted is under dense.}
\label{fig:epxhist}
\end{figure}

In this section we will review and further discuss the relation between the global coordinates and local coordinates that describe the effective curved Universe in presence of a long wavelength mode $\deltal$.
We will be concerned with quantities in the local and global frame, which will
be denoted by subscripts $_\text{L}$ and $_\text{G}$. Furthermore, we can
express local quantities in terms of global and local comoving coordinates and will
denote the quantities measured in terms of global and local coordinates by
superscripts $^\text{(G)}$ and $^\text{(L)}$, respectively.\\
As we have derived in \cite{Baldauf:2011ew}, the effective curved Universe corresponding to a long wavelength, comoving gauge, isotropic, density fluctuation $\delta_\text{l}$ is expanding slightly slower or faster than the homogeneous background Universe leading to a difference in the expansion factors
\be\label{eq:scale_mapping}
a_\text{L}=a_\text{G}\left(1-\frac{\delta_\text{l}}{3}\right).
\ee
Here and in the rest of the paper we limit ourselves to effects linear in $\deltal$, the generalization to non-linear order being straightforward.
Thus, the local expansion rate can be related to the global expansion rate as
\be
H_\text{L}^\text{(G)}(a_\text{G})=\frac{\dot a_\text{L}}{a_\text{L}}=H_\text{G}(a_\text{G})\left(1-\frac{f \delta_\text{l}}{3}\right)\; ,
\ee
where $f=\derd \ln D/\derd \ln a$ is the logarithmic growth factor (see Sec.~\ref{sec:growth} below) and the superscript G is used to show that the local Hubble rate is expressed in terms of the global expansion factor rather than the local one. We will derive the difference between the local expansion rate in global and local coordinates in more detail in Sec.~\ref{sec:hubblerate}. Note that all the expressions in this study are valid to first order in the long wavelength background perturbation $\delta_\text{l}$.
The evolution is described by the effective local density parameters~\footnote{To make contact with the derivation in \cite{Li:2014ss}, we can use the growth factor $D_a$ normalized to $a$ at early times
\[D_a=\frac{5}{2}\Omega_\text{m,0}H_0^2 H \int \derd a\; (a H)^{-3}
\] It is related to our growth factor $D$ as $D=2D_0 D_a /5 H_0^2 \Omega_\text{m,0}$. We can then use Eq.~\eqref{eq:d0} to relate $D_0$ to the logarithmic growth factor and matter density and finally recover 
\[ \Omega_\text{K,L}=-\frac{5}{3} \Omega_\text{m,0}\frac{\delta}{D_a},\]
which agrees with Eq.~(43) in \cite{Li:2014ss}.
}
\be
\begin{split}
\Omega_\text{m,L}=&\frac{8\pi G
\bar{\rho}_\text{L}}{3H_\text{L}^2}=\omnot\left[1
+\left(\omnot+\frac{2}{3}f_0\right)\delta_\text{l,0}\right]\ , \\
\Omega_\text{K,L}=&-\frac{K}{a^2 H^2}=-\left(\omnot+\frac{2}{3}
f_0\right)\delta_\text{l,0}\ , \\
\Omega_{\Lambda,\text{L}}=&\frac{\Lambda}{3H_\text{L}^2}
=\left(1-\omnot\right)\left [
1+\left(\omnot+\frac{2}{3}f_0\right)\delta_\text{l,0}\right]\ .
\end{split}
\ee
valid at $a_\text{L}=1$, where $\delta_\text{l,0}$ is the present day long wavelength density fluctuation and $f_0$ is the logarithmic growth factor in the background Universe at $a_\text{G}=1$.  Since the time at which the local expansion is unity differs from the time at which the global expansion is unity, the above density parameters correspond to $a_\text{G}=(1+{\delta_\text{l,0}}/{3})$.
If the long mode is well within the horizon $k_\text{l}\ll H$, then the gauge
issue is irrelevant and $\delta_\text{l}$ corresponds to the usual Newtonian
overdensity.
The local Hubble rate at $a_\text{L}=1$ can be expressed in terms of the background Hubble $H_\text{G}(a_\text{G}=1)=100\; h\;  \text{km}/\text{s}/\text{Mpc}$
\be
h_\text{L}=h\left[1-\left(\frac{\Omega_\text{m,0}}{2}+\frac{f_0}{3}
\right)\delta_\text{l,0}\right]
\ee
Note that from the above follows
$\left.\Omega_\text{m,L}h_\text{L}^2\right|_{a_\text{L}=1}=\left.\Omega_\text{m,G}h_\text{G}^2\right|_{a_\text{G}=1}$. 
The time dependence of the local density parameters and the local expansion rate are shown in Fig.~\ref{fig:epxhist}.

%===============================================================================
%===============================================================================
\section{How to run Simulations in a curved background}\label{sec:howtorun}
%===============================================================================
\subsection{Growth of Structure}\label{sec:growth}
We are using the linear growth normalised to unity at present time
\be
D(a)=D_0 H(a) \int_0^a \frac{\derd a'}{\bigl[a' H(a')\bigr]^3}
\label{eq:growth}
\ee
where $D_0$ is a normalization factor.

The logarithmic growth factor is thus given by
\begin{align}
f=&\frac{\derd \ln D}{\derd \ln
a}=-\frac{3}{2}\left(1-\Omega_{\Lambda,0}\frac{H_0^2}{H^2}\right)+\frac{D_0}{
H^2a^2 D}=-\frac{3}{2}\frac{\omnot}{a^3}\frac{H_0^2}{H^2}+\frac{D_0}{H^2 a^2 D}
\label{eq:loggrowth}
\end{align}
evaluating the above equation at $a=1$, we have
\be
D_0=H_0^2 \left(f_0+\frac32 \Omega_\text{m,0}\right).
\label{eq:d0}
\ee
%===============================================================================
\subsection{Equivalence of Hubble rate as a function of global and local scale
factor}\label{sec:hubblerate}
Locally, the overdense patch is just a curved Universe and thus the local observer can write down Friedmann equations in terms of his density parameters.
The local Hubble rate as a function of the local scale factor can be written in terms of the local density parameters as
\be
\begin{split}
H^\text{(L)}_\text{L}(a_\text{L})=&h_\text{L}H_{100}\sqrt{\Omega_\text{m,L}
a_\text { L } ^ {
-3 }
+\Omega_\text{K,L}a_\text{L}^{-2}+\Omega_{\Lambda,\text{L}}}\\
=&H_\text{G}(a_\text{L})\left[1-\left(\frac{\Omega_\text{m,0}}{2}
+\frac{f_0
}{3}\right)\frac{\delta_\text{l,0}a_\text{L}}{\Omega_{\Lambda,0}a_\text{L}
^3+\Omega_\text{m,0}
}
\right]
\label{eq:lochubble}
\end{split}
\ee
where
$H_\text{G}(a_\text{L})=h_\text{G}
H_{100}\sqrt{\Omega_\text{m,0}a_\text{L}^{-3}+\Omega_{\Lambda,
0}}$ and we used the mapping between global and local density parameters.
Plugging $a_\text{L}=a_\text{G}\left(1-\frac{\delta_\text{l}}{3}\right)$ into
the
above expression, we obtain the local Hubble rate as a function of the global
scale factor
\be
H^\text{(G)}_\text{L}(a_\text{G})=H_\text{G}(a_\text{G})\left[1+\frac{H_0^2}{H^2
} \left(\frac{\delta_\text{l,0}D \Omega_\text{m,0}}{2a_\text{G}^3}
-\frac{\delta_\text{l,0}\Omega_\text{m,0}}{2a_\text{G}^2}-\frac{\delta_\text{l,0
} f_0 } { 3a_\text{G}^2 } \right) \right]\label{eq:hlrepag}
\ee
Using Eq.~\eqref{eq:loggrowth} we can show that Eq.~\eqref{eq:hlrepag} is equivalent to the previously derived
expression for the local Hubble rate as a function of the global scale factor
\be
H^\text{(G)}_\text{L}(a_\text{G})=H_\text{G}(a_\text{G})\left(1-\frac{f \delta_\text{l}}{3}\right).
\ee
%===============================================================================
\subsection{Mapping of time}

The local closed Friedmann-Lema\^{i}tre-Robertson-Walker (FLRW) spacetime that we have defined in~\cite{Baldauf:2011ew} should be thought of as a chart covering a sub-Hubble patch of the perturbed FLRW manifold. In~\cite{Baldauf:2011ew} we provided the change of coordinates between the local and the global frame, so that we know exactly how to identify a point on the manifold expressed in local coordinates with the global ones. Here this is expressed in~Eq.~(\ref{eq:scale_mapping}). We can however gain some intuition on the mapping of the coordinates by realizing the following. Both in the closed FLRW and in the perturbed FLRW, time delay is a general relativistic effect that is negligible. This implies that for the mapping of coordinates given by~Eq.~(\ref{eq:scale_mapping}) should correspond to having the same FLRW time both in the closed patch and in the global patch. Let us check this. \\
The time in the effective curved patch is given by
\be
\begin{split}
t=&\int_0^{a^*_\text{L}} \frac{\derd a'}{H_\text{L}(a')a'}\\
=&\int_0^{a_\text{G}}\frac{\derd a'}{H_\text{G}(a')\,
a'}\left[1+\left(\frac{\Omega_\text{m,0}}{2}+\frac{f_0
}{3}\right)\frac{\delta_\text{l,0}a'}{\Omega_{\Lambda,0}a^3+\Omega_\text{m,0}}
\right] +\frac{a_\text{L}-a_\text{G}}{H_\text{G}
(a_\text{G})a_\text{G} }
\end{split}
\ee
Here we used Eq.~\eqref{eq:lochubble} and the fact that the difference between local and global expansion factor
corresponding to the same proper time is a first order quantity.
We want the proper time in the curved frame to agree with the proper time in
the fiducial flat background
\begin{equation}
\begin{split}
a_\text{L}=&a_\text{G}\left[1-H_\text{G}\int_0^{a_\text{G}}\frac{\derd
a'}{H_\text{G}(a')}\left(\frac{\Omega_\text{m,0}}{2}+\frac{f_0
}{3}\right)\frac{\delta_\text{l,0}}{\Omega_{\Lambda,0}{a'}^3+\Omega_\text{m,0}}
\right]\\
=&a_\text{G}\left[1-\sqrt{\Omega_\text{m}
a^{-3}+\Omega_\Lambda}\left(\frac{\Omega_\text{m,0}}{2}+\frac{f_0
}{3}\right)\delta_\text{l,0}
\int_0^{a_\text{G}}\frac{\derd
a'}{\left(\Omega_{\Lambda,0}{a'}
^2+\Omega_\text{m,0}{ a'}^{-1}\right)^{3/2}}
\right]\\
=&a_\text{G}\left[1-\frac{D}{D_0}H_0^2 \left(f_0+\frac32 \Omega_\text{m}\right)\frac{\delta_\text{l,0}}{3}\right]\\
=&a_\text{G}\left[1-\frac{\delta_\text{l}}{3}\right]
\end{split}
\end{equation}
This little calculation confirms the na\"{\i}ve expectation that $a_\text{L}=a_\text{G}(1-{\delta_\text{l}}/{3})$ and thus Eq.~\eqref{eq:scale_mapping}.
%===============================================================================
\subsection{Initial Conditions and Units}
At early times the amplitude of the long mode is
suppressed and thus the mean densities in the local and global patch agree.
As customary in the LSS community, we will express local lengths and masses in terms of
$h_\text{L}^{-1}\text{Mpc}$ and $h_\text{L}^{-1}M_\odot$. This means that in the curved Universe simulations the length and mass unit is different from the flat case and consequently the numerical prefactor has to be changed accordingly. For example, for the box size we have
\be
L_\text{G}= 500\ h_\text{G}^{-1} \text{Mpc}=500\frac{h_\text{L}}{h_\text{G}}\ h_\text{L}^{-1} \text{Mpc}=L_\text{L}.
\ee
The global critical density is given by
\be
\rho_\text{crit,G}=\frac{3 H_0^2}{8 \pi G}=27.75
\tim{10} h_\text{G}^2 \frac{M_\odot}{\Mpc^3}
\ee
and for the local one we have to replace $h_\text{G}$ by $h_\text{L}$.
The particle mass is given by
\be
M_\text{p,G}=\frac{L_\text{G}^3
\rho_\text{crit,G}\Omega_\text{m,G}}{N_\text{p}^3}=13.875\tim{13} \frac{\Omega_\text{m,G}}{N_\text{p}^3} \ h_\text{G}^{-1}M_\odot.
\ee
where $N_\text{p}$ is the number of particles per dimension. Thus the units of the particle mass are $[M_\text{p}]=h^{-1} M_\odot$.
The local particle mass is given by
\be
\begin{split}
M_\text{p,L}=\frac{L_\text{L}^3
\rho_\text{crit,L}\Omega_\text{m,L}}{N_\text{p}^3}=&13.875\tim{13} \frac{h_\text{L}^3}{h_\text{G}^3}\frac{\Omega_\text{m,L}}{N_\text{p}^3} \ h_\text{L}^{-1}M_\odot\\
=&13.875\tim{13} \frac{h_\text{L}}{h_\text{G}}\frac{\Omega_\text{m,G}}{N_\text{p}^3} \ h_\text{L}^{-1}M_\odot=M_\text{p,G}
\end{split}
\ee
Here we used that $\Omega_\text{m,L}h_\text{L}^2=\Omega_\text{m,G}h_\text{G}^2$.
Thus, global and local particle mass agree up to the $h$ remapping. The change of mass and length units doesn't affect any results, only the numerical factors entering into the simulation codes. When extracting measurements from the simulations we convert to the global length unit and all the results given in this paper are in terms of the global length and mass units $h_\text{G}^{-1}\text{Mpc}$ and $h_\text{G}^{-1}M_\odot$.

%===============================================================================
\subsection{Rescaling in the End}\label{sec:rescend}
Using that the local and global patch correspond to the same physical scale
$r=a_\text{G} x_\text{G}=a_\text{L} x_\text{L}$, we can map from local comoving coordinates to global comoving coordinates as
\be
x_\text{G}(x_\text{L})
=x_\text{L}
\left(1-\frac{ \delta_\text{l}}{3} \right)\; .
\label{eq:scaleshift}
\ee
At very early times, before the long mode could influence the expansion, the scales of comoving features (for instance the BAO) are equal $x_\text{G}=x_\text{L}$.
This basically means that if a correlation function in the local patch, $\xi_\text{L}$, has a feature in the initial conditions, such as the BAO peak, at $x_\text{L}=x_\text{L}^\text{F}$, the global coordinate corresponding to this feature will be $x_\text{G}^\text{F}=x_\text{L}^F(1+\deltal/3)$.
For the overdense Universe simulation box, this also means that the globally observed length of the whole box is reduced.
This rescaling makes sure that
$\bar \rho_\text{L}=\bar \rho_\text{G}(1+\delta_\text{l})$ in global
coordinates since
\be
\bar \rho_\text{L}=\frac{N_\text{p} M_\text{p,L}}{[L_\text{L}^{(\text{G})}]^3}=\frac{N_\text{p} M_\text{p,L}}{[L_\text{L}^{(\text{L})}]^3}(1+\delta_\text{l})=\bar \rho_\text{G}(1+\delta_\text{l})
\ee
since we have chosen the local comoving box scale to agree with the global one $L_\text{L}^{(\text{L})}=L_\text{G}$. Note however, that the comoving density at $a_\text{L}=1$ agrees with the global one at $a_\text{G}=1$, since
\be
\bar \rho_\text{L,0}=\rho_\text{crit,L} \Omega_\text{m,L}=\rho_\text{crit,G} \frac{h_\text{L}^2}{h_\text{G}^2} \Omega_\text{m,L}=\rho_\text{crit,G} \Omega_\text{m,G}=\bar \rho_\text{G,0}.
\ee

%===============================================================================

\subsection{The Simulations}
The goal of this study is to numerically investigate the effect of long wavelength fluctuations on local dynamics. As mentioned before, for spherical configurations, this corresponds to measure the corresponding quantity in a slightly over- or underdense patch and to infer the derivative numerically. For this purpose we employ a fiducial flat simulation $\delta_\text{l}=0$ and symmetric over- and underdense simulations with $\delta_\text{l,+}=0.1$ and $\delta_\text{l,-}=-0.1$.\\
The initial conditions are set up at redshift $z_\text{i}=99$. The transfer functions are calculated using
\texttt{CMBFAST} \cite{Seljak:1996ap}, the simulations with different curvature parameters share the same
realization of modes in the box to cancel cosmic variance.
The gravitational evolution of the dark matter particles is followed using the publicly available $N$-body code \texttt{Gadget II}~\cite{Springel:2005mi}. 
The box sizes are S: $L=80 \hMpc$, M: $L=500 \hMpc$ and L: $L=1600 \hMpc$ and we use $N=512^3$ (S, M) and $N=1024^3$ (L) particles to sample the matter density field. 
The minimum halo masses are $M_\text{h,min,S}=5.9\tim{9}\hMs$ $M_\text{h,min,M}=1.4\tim{12}\hMs$ $M_\text{h,min,L}=5.6\tim{12}\hMs$, for the S, M and L simulations respectively.
The fiducial parameters are inspired by the WMAP 5-year analysis~\cite{Komatsu:2008hk} and the corresponding parameters for the curvature simulations are given in Table \ref{tab:cosmoparam}. The fluctuations of the fiducial and variant curvature simulations are normalised at the initial redshift to the amplitude $A=2.21\tim{-9}$ at the pivot scale $k_\text{pivot}=0.02\ \text{Mpc}^{-1}$. We also run a set of simulations with zero curvature but variant amplitude, whose values are given in Table~\ref{tab:sigmatab}.\\
Bound objects are identified using a Friends-of-Friends (FoF) halo finder with a linking length of $l=0.2$ mean inter particle spacings of the fiducial simulation. Since the mean interparticle spacing is increased (reduced) in the overdense (underdense) box, we need to adjust the linking length accordingly.
Thus, when finding the haloes we should either use the halo finder with
the modified linking length $l(1+\delta_\text{l}/3)$ in local comoving coordinates or
rescale the simulation coordinates to global coordinates and then use the linking length of the flat
simulation. We employed the first of these options.\footnote{There are some additional short distance parameters in \texttt{GADGET} that depend on $h$, such as the force smoothing scale. We do not rescale these parameters, as we expect that they do not affect the result. Indeed, since they are not physical, if they were to affect the result, it would be that the result is systematically incorrect.}
%>>>>>>>>>>>>>>>>>>>>>>>>>>>>>>>>>>>>>>>>>>>>>>>>>>>>>>>>>>>>>>>>>>>>>>>>>>>>>>>>>>>
\begin{table}[htbp]
\centering
\begin{tabular}{lrrr}
\hline
\hline
 & - & f& +\\
\hline
$\delta_\text{l}$ & -0.100 & 0.00 & 0.100 \\
$\Omega_\text{m}$ & 0.263 & 0.28 & 0.297 \\
$\Omega_\text{b}$ & 0.038 & 0.04 & 0.042 \\
$\Omega_\text{c}$ & 0.225 & 0.24 & 0.255 \\
$\Omega_\text{K}$ & 0.061 & 0. & -0.061 \\
$\Omega_\Lambda$ & 0.676 & 0.72 & 0.764 \\
$a_\text{L}(a_\text{G}=1)$ & 1.033 & 1. & 0.967 \\
 \text{h} & 0.721 & 0.7 & 0.679 \\
\hline
\hline
\end{tabular}
\caption{Cosmological parameters of the simulations.}
\label{tab:cosmoparam}
\end{table}
%>>>>>>>>>>>>>>>>>>>>>>>>>>>>>>>>>>>>>>>>>>>>>>>>>>>>>>>>>>>>>>>>>>>>>>>>>>>>>>>>>>>
%
%>>>>>>>>>>>>>>>>>>>>>>>>>>>>>>>>>>>>>>>>>>>>>>>>>>>>>>>>>>>>>>>>>>>>>>>>>>>>>>>>>>>
\begin{table}[h]
\begin{center}
\begin{tabular}{llll}
\hline
\hline
&LS2 & FID & HS2\\
A &
$1.989\tim{-9}$&
$2.21\tim{-9}$&
$2.43\tim{-9}$\\
$\sigma_8$ & 0.729 & 0.81 & 0.891 \\
\hline
\hline
\end{tabular}
\end{center}
\caption{Scalar amplitudes for the variant $\sigma$ simulations.}
\label{tab:sigmatab}
\end{table}
%>>>>>>>>>>>>>>>>>>>>>>>>>>>>>>>>>>>>>>>>>>>>>>>>>>>>>>>>>>>>>>>>>>>>>>>>>>>>>>>>>>>

%===============================================================================
%===============================================================================
\section{Overdensities and Power Spectrum}\label{sec:power}
The goal of this section is to relate what is seen by a global observer in a local overdense patch to quantities computable in the background Universe. In particular, we want to relate the fluctuation spectrum in the locally overdense patch to the fluctuation spectrum of the background cosmology. This allows us to measure the physical squeezed limit~\cite{Creminelli:2013cga} of three point functions (or in general $n+1$ point functions) by measuring the power spectrum (or equivalently $n-$point functions) in curved simulations. We will proceed in three steps: first we will discuss that locally measured overdensities need to be rescaled in order to correspond to global overdensities. Then we will discuss the change of the linear growth in the overdense patch and finally we will discuss how the change in expansion rate affects the scale at which the background fluctuation spectrum needs to be evaluated to describe the local comoving features. A sketch describing the local and global coordinates and overdensities is given in Fig.~\ref{fig:skecth}.
\subsection{Mean Density}
\begin{figure}
\centering
\includegraphics[width=0.99\textwidth]{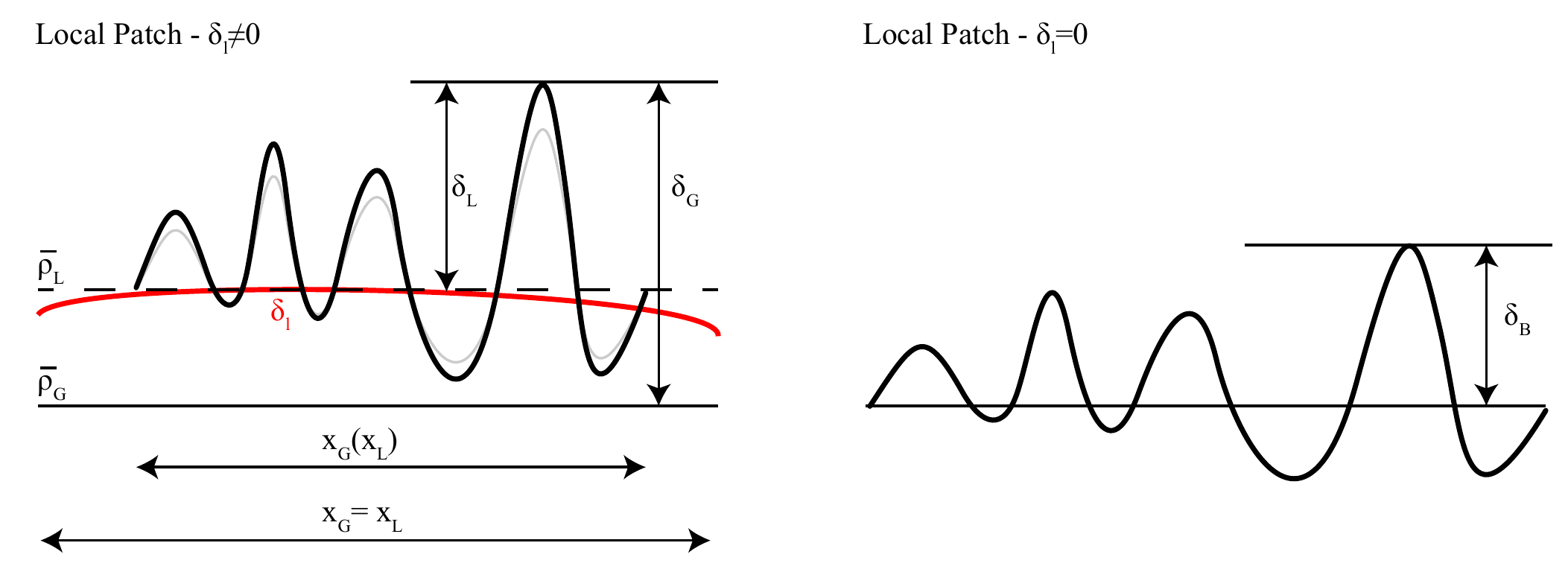}
\caption{Sketch of a local patch with and without the long mode $\deltal$ (considered to be an overdensity here). If the mode is present then the locally measured overdensities of short fluctuations $\delta_\text{L}$, need to be rescaled to correspond to a globally normalized overdensity $\delta_\text{G}$. The growth of structures is enhanced in the local patch and the observed size of an evolved comoving feature appears smaller due to the reduced expansion.}
\label{fig:skecth}
\end{figure}

If we observe some overdensity from a global perspective, we measure the overdensity with respect to the global mean
\be
\delta_\text{G}(\vec x)=\frac{\rho(\vec x)}{\bar{\rho}_\text{G}}-1.
\ee
From a local perspective, the same overdensity is assessed with respect to the local mean density $\bar \rho_\text{L}$ as (see Fig.~\ref{fig:skecth})
\be
\delta_\text{L}(\vec x)=\frac{\rho(\vec x)}{\bar{\rho}_\text{L}}-1 \; ,
\ee
where we evaluated the density at a fixed global coordinate spacetime point $\vec{x}$.
We can now convert the background density to the global density
\be
\begin{split}
\delta_\text{L}(\vec x)=&\frac{\rho(\vec x)}{\bar{\rho}_\text{G}(1+\delta_\text{l})}-1\\
=&\frac{1+\delta_\text{G}(\vec x)}{1+\delta_\text{l}}-1,
\end{split}
\ee
thus each locally measured overdensity has to be rescaled as
\be
\delta_\text{G}(\vec x)=\delta_\text{L}(\vec x)(1+\delta_\text{l})+\delta_\text{l}.
\label{eq:meandens}
\ee
to find its equivalent as seen by a global observer.
The additional additive $\delta_\text{l}$ is constant over the box and does thus not correlate with the fluctuations within the box.
%%%%%%%%%%%%%%%%%%%%%%%%%%%%%%%%%%%%%%%%%%%%%%%%%%
\subsection{Growth in Presence of the Background Mode}
The linear evolution of fluctuations in presence of a background mode can be assessed by the linear growth factor Eq.~\eqref{eq:growth}, evaluated in the effective locally curved Universe. Compared to the matter only background Universe the growth is changed by the explicit dependence of the growth factor on the effective curvature at fixed local expansion and also by the change of the local expansion factor. In a matter only Einstein--de Sitter (EdS) background, we have $D_\text{G}(a_\text{G})\propto a_\text{G}$ and thus
\be
\left.\frac{\partial D}{\partial \delta_\text{l}}\right|_{a_\text{G}}=\frac{\partial D}{\partial \deltal}\Biggr|_{a_\text{L}}+\frac{\partial D}{\partial a}\Biggr|_{\deltal}\frac{\partial a}{\partial \deltal}=\frac{20}{21}a_\text{G}-\frac{1}{3}a_\text{G}=\frac{13}{21}a_\text{G}\ .
\label{eq:growthloceds}
\ee
This means that the linear growth of short modes within the patch is enhanced with respect to modes that live in a patch without a long mode as $D_\text{L}=a_\text{L}(1+13/21 \deltal)$.
Let us now consider this relation for a more general $\Lambda$CDM background Universe
\be
\begin{split}
D_\text{L}(a_\text{L})=&\frac{5}{2}\Omega_\text{m,L} H_{0,\text{L}}^2 H_\text{L}(a_L) \int_0^{a_\text{L}}\frac{\derd a'}{[a' H_\text{L}(a')]^3} \\
=&\frac{5}{2}\Omega_\text{m,G} H_{0,\text{G}}^2 H_\text{L}(a_L) \int_0^{a_\text{L}} \frac{\derd a'}{[a' H_\text{L}(a')]^3}\ .
\end{split}
\ee
We want this quantity to first order in the long wavelength perturbation, evaluated at the local expansion factor corresponding to a certain global expansion.
\be
\begin{split}
D_\text{L}(a_\text{G})=&\frac{5}{2}\Omega_\text{m,G} H_{0,\text{G}}^2 H_\text{G}(a_G)\left(1-\frac{f \delta_\text{l}}{3}\right)
\int_0^{a_\text{G}\left(1-\frac{{ \deltal}}{3}\right)} \frac{\derd a'}{[a' H_\text{G}(a')]^3} \left[1+3\left(\frac{\Omega_\text{m,0}}{2}+\frac{f_0
}{3}\right)\frac{\delta_\text{l,0}a'}{\Omega_{\Lambda,0}a^3+\Omega_\text{m,0}}
\right]\\
=&D_\text{G}(a_\text{G})+\frac{5}{2} \frac{\Omega_\text{m,G} H_{0,\text{G}}^2}{[a_\text{G}H_\text{G}(a_\text{G})]^2}\left(-\frac{\delta_\text{l}}{3}\right)-\frac{f\deltal}{3}D_\text{G}(a_\text{G})
+\frac{5}{2}\Omega_\text{m,G}H_{0,G}^2H_\text{G}(a_\text{G}) D_0 \delta_{\text{l},0}\int \frac{\derd a'}{[a' H_\text{G}(a')]^5}\\
=&D_\text{G}(a_\text{G})\left(1-\frac{f\delta_\text{l}}{3}\right)-\frac52 \frac{\delta_\text{l}}{3}\frac{\Omega_\text{m,G} H_{0,\text{G}}^2}{[a_\text{G}H_\text{G}(a_G)]^2}
+\frac{5}{2}\Omega_\text{m,G}H_{0,G}^2\delta_{\text{l},0}D_0 H_\text{G}(a_G)\int \frac{\derd a'}{[a' H_\text{G}(a')]^5}\ .
\end{split}
\ee
For an EdS background the above equation yields $a_\text{G}(1+13/21 \delta_\text{l})$ and deviates at the sub-percent level for $\Lambda$CDM.\footnote{For the cosmology under consideration here, the relative deviation of the local growth from the EdS case is $+0.42\%$. This deviation is in perfect agreement with the exact growth factors for standard perturbation theory \cite{Bernardeau:2002} in $\Lambda$CDM \cite{Takahashi:2008yk}.}
%>>>>>>>>>>>>>>>>>>>>>>>>>>>>>>>>>>>>>>>>>>>>>>>>>>>>>>>>>>>>>>>>>>>>>>>>>>>>>>>>>>
%<<<<<<<<<<<<<<<<<<<<<<<<<<<<<<<<<<<<<<<<<<<<<<<<<<<<<<<<<<<<<<<<<<<<<<<<<<<<<<<<<<

\subsection{Shift and Power Spectrum}\label{sec:shift}

The former derivation allows us to write the locally measured overdensity as
\be
\delta_\text{L}(\vec x)=\left(1+\frac{13}{21}\deltal\right) \delta_\text{B}\left[\vec x\left(1+\frac{\deltal}{3}\right)\right]\; ,
\ee
where $\delta_\text{B}(x)$ is the linearly evolved overdensity for the background cosmology ($\Omega_K=0$, $\deltal=0$). Using~Eq.~\eqref{eq:meandens}, and the shift in coordinates from the local to the global frame~(\ref{eq:scaleshift}), we can obtain an expression for the globally normalized overdensity evolved at linear order in the short modes and at linear order in the effect of the long mode: 
\be
\delta_\text{G}(\vec x)=(1+\deltal)\;\delta_\text{L}\left(\vec x\right)=\left(1+\frac{13}{21}\deltal\right)\,(1+\deltal)\;\delta_\text{B}\left[\vec x\left(1+\frac{\deltal}{3}\right)\right]\; .
\label{eq:longmodeguess}
\ee
Thus the local correlation function can be expressed in terms of the background correlation function $\xi_\text{B}$ as
\be
\label{eq:xiloc}
\xi_\text{G}(r)=\left(1+\frac{68}{21}\delta_\llg\right)\xi_\text{B}\left[r\left(1+\frac{\delta_\llg}{3}\right)\right]\; .
\ee
Likewise, wavenumbers in Fourier space can be expressed in terms of local or global comoving coordinates. In particular the locally observed wavenumber of a comoving feature is now enhanced in an overdense patch
\be
k_\text{G}(k_\text{L})=k_\text{L}\left(1+\frac{\delta_\text{l}}{3}\right).
\ee
This means that the power at a certain wavenumber expressed in local coordinates
corresponds to the power at a rescaled wavenumber in global coordinates 
(see also \cite{Sherwin:2012ar}). 
The Fourier grid that shared the same seeds and comoving spacing initially is now expanded or contracted. Thus a mode that shared the same seed as the global mode $k_\text{G}$ in the initial conditions (i.e., $k_\text{L}=k_\text{G}$) is now observed at $k_\text{G}(k_\text{L})=k_\text{L}(1+\delta_\text{l}/3)$. 
Since we have $k_\text{L}=k_\text{G}(k_\text{L})$ initially, by comparing modes at $k_\text{L}$ in the flat Universe with modes
at $k_\text{G}(k_\text{L})=k_\text{L}(1+\delta_\text{l}/3)$ in the curved Universe, we can cancel the cosmic variance.\\
When extracting statistics from our curvature simulations we first map the local comoving coordinates to the global comoving coordinates corresponding to the same time as described above in Sec.~\ref{sec:rescend}. This mimics the experience of an external observer. Consequently, the measured power spectrum is given in terms of $k_\text{G}(k_\text{L}) $. To asses the change in growth we thus have to evaluate the following ratio
\be
w^2=\frac{P_\text{G}\left[k_\text{L}\left(1+\frac{\deltal}{3}\right)\right]}{P_\text{B}(k_\text{L})}.
\ee
This ratio is given by the enhanced growth in the local coordinates Eq.~\eqref{eq:growthloceds} and the mean density rescaling according to Eq.~\eqref{eq:meandens} and finally a factor $(1-\delta_\text{l})$ arises from the fact that the power spectrum has units of volume\footnote{For a finite volume (like a simulation box) we have $P(k_i)=1/V_\text{G}(V_\text{L})\ \la \delta(\vec k_i) \delta(-\vec k_i)\ra$ and $\delta(\vec k_i)\propto V_\text{G}(V_{\text{L}})$, thus $P\propto V_\text{G}(V_\text{L})= V_\text{L}(1-\deltal)$.}
\be\label{eq:powerenhancefac}
w^2=\left(1+ \frac{13}{21} \delta_\text{l}\right)^2\left(1+\delta_\text{l}\right)^2\left(1-\delta_\text{l}\right)\approx1+\frac{47}{21}\delta_\text{l}\; .
\ee
The shift in the argument of the local power spectrum can be expressed as 
\begin{align}
P_\text{G}\left[k_\text{L}\left(1+\frac{\delta_\text{l}}{3}
\right)\right]=& P_\text{G}(k_\text{L})+k_\text{L}
P_\text{G}'(k_\text{L})\frac{\delta_\text{l}}{3}
=P_\text{G}\left(k_\text{L}\right)\left[1+\frac{\delta_\text{l}}{3} 
\frac{\derd \ln P}{\derd \ln k}\right].
\end{align}
If we would like to compare the locally observed power spectrum to the background for the same wavenumber argument we obtain
\be
\frac{P_\text{G}(k_\text{L})}{P_\text{B}(k_\text{L})}=w^2\left[1-\frac{\delta_\text{l}}{3} 
\frac{\derd \ln P}{\derd \ln k}\right],
\label{eq:addlgderivp}
\ee
i.e., we need to correct for the change of the power spectrum due to the local contraction/expansion.
The growth enhancement can also be seen from Standard Perturbation Theory \cite{Bernardeau:2002}, where the effect of a spherically symmetric long wavelength mode $\delta_\text{l}(\left|\vec q\right|)=2\pi^2\ddir(q-q_0)q^{-2} \delta_\text{l}$ on a short wavelength mode $\delta(\vec k)$ with $k\gg q$ can be estimated from the coupling of a linear background mode $\delta_\text{B}^{(1)}$ and the long mode
\begin{equation}
\begin{split}\label{eq:localdenskspace}
\delta_\text{G}(\vec k)=&\delta_\text{B}^{(1)}(\vec k)+2\int \dqc F_2(\vec q,\vec k-\vec q)\delta_\text{l}^{(1)}(\vec q)\delta_\text{B}^{(1)}(\vec k-\vec q)+\mathcal{O}(\delta^3)\\
\approx&\delta_\text{B}^{(1)}(\vec k)+2\int \dqc F_2(\vec q,\vec k-\vec q)\delta_\text{l}(|\vec q|)\Bigl[\delta_\text{B}^{(1)}(\vec k)-\vec q\cdot \vec \nabla\delta_\text{B}^{(1)}(\vec k)\Bigr]\\
\approx&\delta_\text{B}^{(1)}(\vec k)+2\int \dqc \left(\frac{3}{14}+\frac12 \frac{\vec k\cdot \vec q}{q^2}+\frac27 \frac{(\vec k\cdot \vec q)^2}{k^2 q^2}\right)\delta_\text{l}(|\vec q|)\Bigl[\delta_\text{B}^{(1)}(\vec k)-\frac{\vec q\cdot \vec k}{k^2} \vec k\cdot \vec \nabla\delta_\text{B}^{(1)}(\vec k)\Bigr]\\
\approx& \delta_\text{B}^{(1)}(\vec k)+\frac{13}{21}\delta_\text{l}\delta_\text{B}^{(1)}(\vec k)-\frac{1}{3}\delta_\text{l}\vec k \cdot \vec \nabla \delta_\text{B}(\vec k) =\delta_\text{B}^{(1)}\left[\vec k\left(1-\frac{\delta_\text{l}}{3}\right)\right]+\frac{13}{21}\delta_\text{B}^{(1)}(\vec k)\delta_\text{l}\ ,
\end{split}
\end{equation}
where in the third line we projected the gradient along the $\vec k$ direction (the components orthogonal to $k$ do not contribute after angular integration).
Note that this expression is automatically normalized to the global mean density.
Equivalently, we can write the same quantity using the real space expressions for the second order density field (see e.g. \cite{Sherwin:2012ar}), the displacement field $\vec \Psi$ and tidal tensor $s_{ij}$, obtaining
\be
\begin{split}\label{eq:localdensrescale}
\delta_\text{G}(\vec x)=&\delta^{(1)}(\vec x)+\frac{17}{21}\left[\delta^{(1)}(\vec x)\right]^2-\vec \Psi(\vec x) \cdot \vec \nabla \delta^{(1)}(\vec x)+s^2(\vec x)\\
=&\delta_\text{B}^{(1)}(\vec x)+\frac{34}{21}\delta_\text{l}\delta_\text{B}^{(1)}(\vec x)-\vec \Psi_\text{l}(\vec x) \cdot \vec \nabla\delta_\text{B}^{(1)}(\vec x)\\
=&\delta_\text{B}^{(1)}(\vec x)+\frac{34}{21}\delta_\text{l}\delta_\text{B}^{(1)}(\vec x)+ \frac13\delta_\text{l} \vec x \cdot \vec \nabla \delta_\text{B}^{(1)}(\vec x)\\
=&\left(1+\frac{34}{21}\delta_\text{l}\right)\delta_\text{B}^{(1)}\left[\vec x\left(1+\frac13 \delta_\text{l}\right)\right]\\
\end{split}
\ee
This result agrees with the heuristically derived result in Eq.~\eqref{eq:longmodeguess}.

\begin{figure}
\centering
\includegraphics[width=0.49\textwidth]{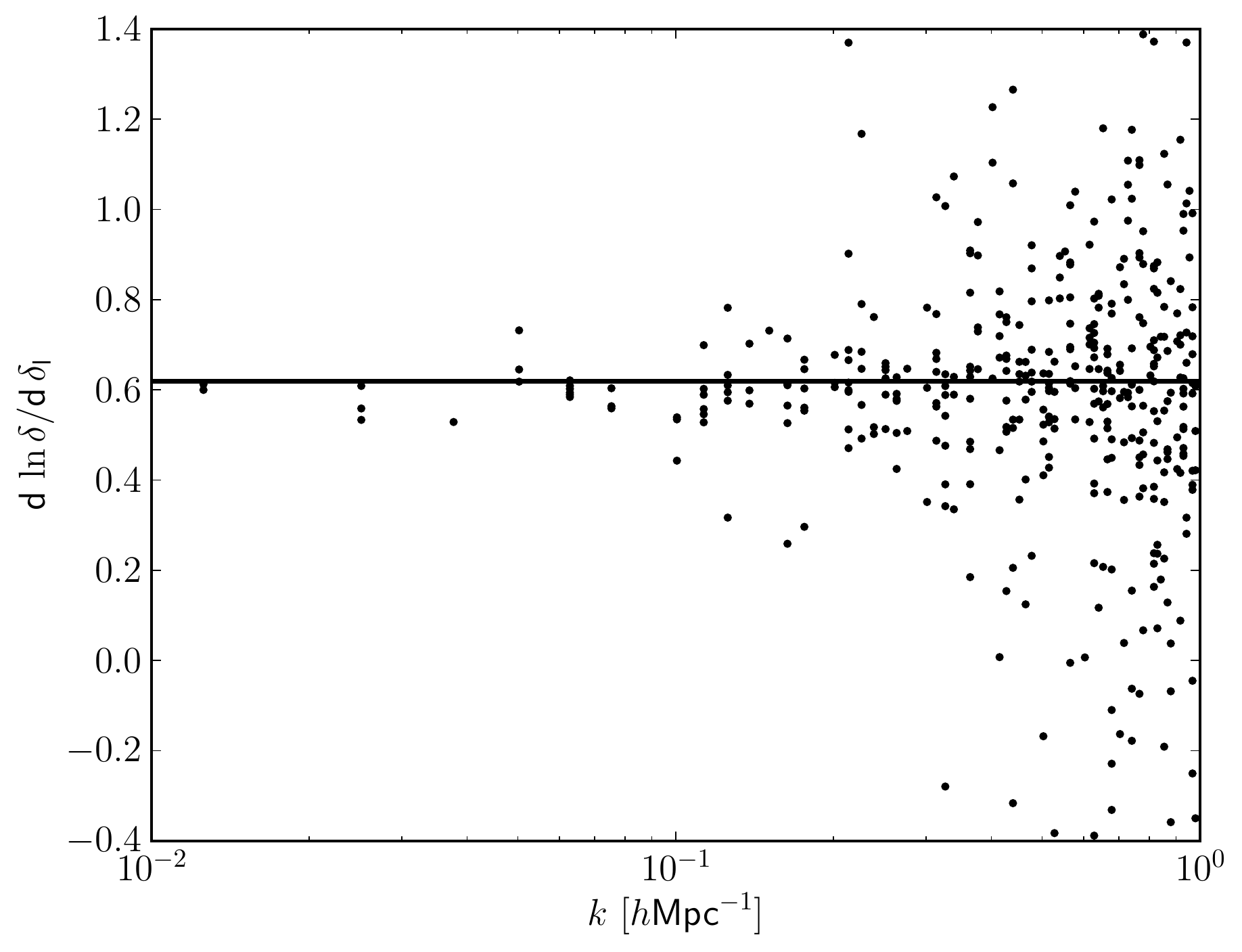}
\caption{Fractional deviation between the final $k$-space modes, where we are focusing on the growth correction $1+13/21 \deltal$. 
We show the symmetric logarithmic derivative, where the quadratic corrections cancel. The three fundamental modes are lying on top of each other, but at higher wavenumbers there is considerable stochasticity around the mean response.}  
\label{fig:modes}
\end{figure}

\begin{figure}
\centering
\includegraphics[width=0.49\textwidth]{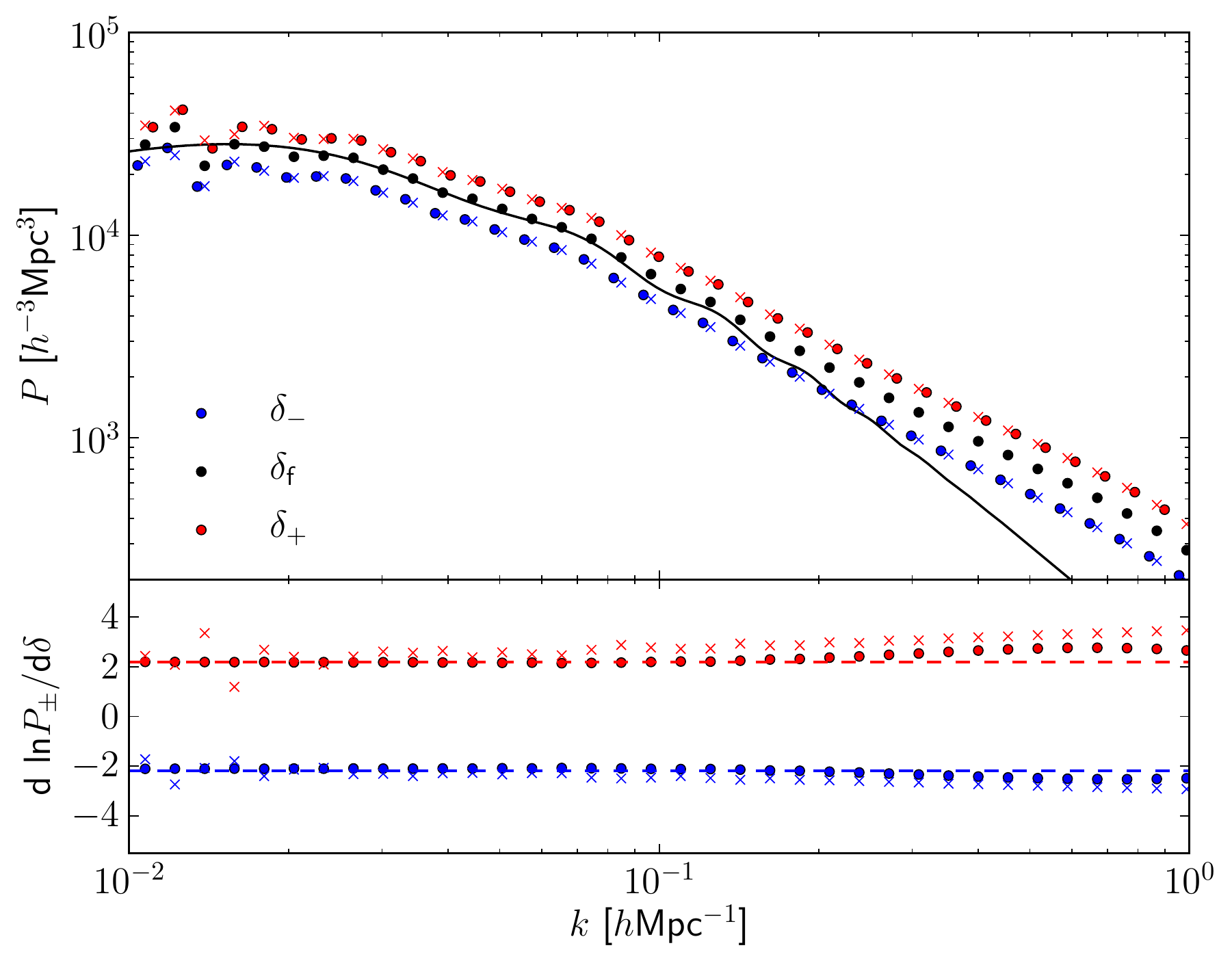}
\includegraphics[width=0.49\textwidth]{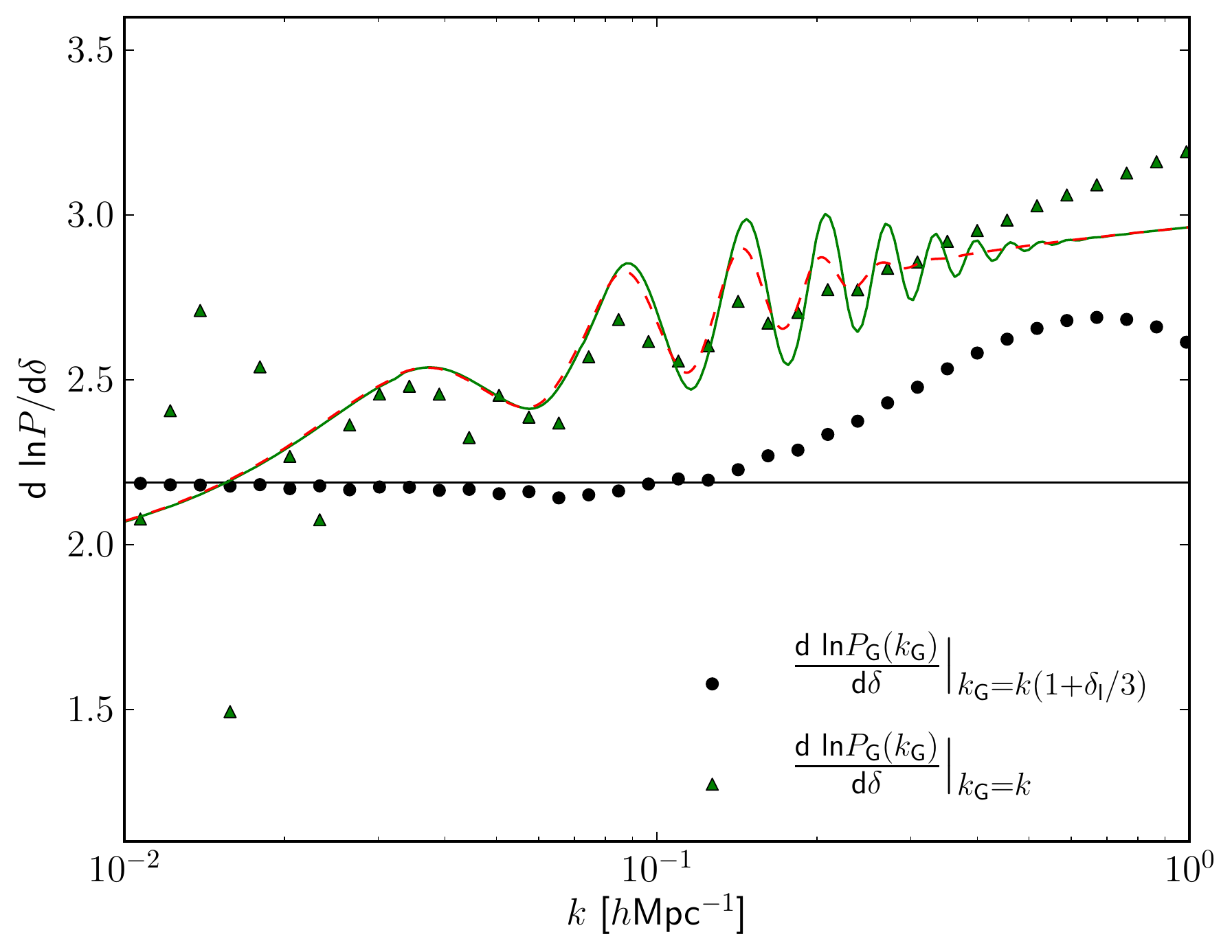}
\vspace{-0.4cm}
\caption{\emph{Left upper panel: }Power spectrum of the fiducial and curved simulations. The dots show the power at the local wavenumber in global coordinates $k_\text{G}(k_\text{L})$, whereas the crosses show the local power interpolated to the global wavenumber. \emph{Left lower panel: } One sided derivatives of the power spectrum to the long mode. The circles describe the growth and mean density effect only, and the crosses take into account the shift in wavenumber as well. \emph{Right panel: } Logarithmic derivative of the power with respect to the long wavelength mode. The dots arise from a comparison of the modes with the same initial seeds, i.e. they compare the local power at $k(1+\deltal/3)$ to the background power at $k$, whereas the triangles also account for the wavenumber shift due to the expansion/contraction. The horizontal solid line shows the $47/21$ growth and mean density and volume enhancement Eq.~\eqref{eq:powerenhancefac}. The green line accounts for the wavenumber shift by adding the logarithmic derivative of the power spectrum according to Eq.~\eqref{eq:addlgderivp} and the red dashed line adds IR-resummation to this result. 
}
\label{fig:power}
\end{figure}

We can recover the Fourier space expression, explicitly performing the transformation
\be
\begin{split}
\delta_\text{G}(\vec k)=&\int \derd^3x\; \delta_\text{G}(\vec x)\eh{i \vec k\cdot \vec x}\\
=&\left(1+\frac{34}{21}\delta_\text{l}\right)\int \derd^3x\; \delta_\text{B}^{(1)}\left[\vec x\left(1+\frac13 \delta_\text{l}\right)\right]\eh{i \vec k\cdot \vec x}\\
=&\left(1+\frac{34}{21}\delta_\text{l}\right)(1-\delta_\text{l})\int \derd^3\tilde x\; \delta_\text{B}^{(1)}(\tilde{\vec x})\eh{i \tilde{\vec x}\cdot \vec k \left(1-\frac{1}{3}\delta_\text{l}\right)}\\
=&\left(1+\frac{13}{21}\delta_\text{l}\right)\delta_\text{B}^{(1)}\left[\vec k\left(1-\frac{\delta_\text{l}}{3}\right)\right]\; ,
\end{split}
\ee
which agrees with Eq.~\eqref{eq:localdenskspace}. We see again that the difference between the $13/21$ and $34/21$ enhancements arises from the measure, i.e., the volume rescaling. The above scaling is confirmed for the modes in the curvature simulations, as can be seen in Fig.~\ref{fig:modes}. On large scales, for instance for the fundamental mode, the response is perfect but on smaller scales (higher wavenumbers) there is considerable scatter around the mean response.
From the above expression for the Fourier modes we can calculate the power spectrum
\begin{equation}
\begin{split}
\la \delta_\text{G}(\vec k)\delta_\text{G}(\vec k')\ra=&P_\text{G}(k)(2\pi)^3\ddir(\vec k+\vec k')\\
=&\left(1+\frac{26}{21}\delta_\text{l}\right)\la \delta_\text{B}^{(1)}\left[\vec k\left(1-\delta_\text{l}/3\right)\right]\delta_\text{B}^{(1)}\left[\vec k'\left(1-\delta_\text{l}/3\right)\right]\ra\\
=&\left(1+\frac{26}{21}\delta_\text{l}\right) (2\pi)^3 \ddir\bigl((\vec k+\vec k')\left(1-\delta_\text{l}/3\right)\bigr)P_\text{B}\left[k\left(1-\frac{\delta_\text{l}}{3}\right)\right]\\
=&\left(1+\frac{26}{21}\delta_\text{l}\right) (2\pi)^3 \frac{\ddir(\vec k+\vec k')}{1-\delta_\text{l}}P_\text{B}\left[k\left(1-\frac{\delta_\text{l}}{3}\right)\right]
\label{eq:deltatop1}
\end{split}
\end{equation}
Thus we finally obtain for the power spectrum
\be
P_\text{G}(k)=\left(1+\frac{47}{21}\delta_\text{l}\right)P_\text{B}\left[k\left(1-\frac{\delta_\text{l}}{3}\right)\right]\; .
\label{eq:deltatop2}
\ee
The same result could have been obtained starting from the correlation function in Eq.~\eqref{eq:xiloc}
\begin{equation}
\begin{split}
P_\text{G}(k)=&\int \derd^3 r\, \xi_\text{G}(r)\eh{\ii \vec k \vec r}\\
=&\left(1+\frac{68}{21}\delta_\llg\right)\int \derd^3 r\, \xi_\text{B}\left[r\left(1+\frac{\delta_\llg}{3}\right)\right]\eh{\ii \vec k \vec r}\\
=&\left(1+\frac{68}{21}\delta_\llg\right)(1-\delta_\llg) \int \derd^3 \tilde{r}\, \xi_\text{B}(\tilde{r}) \eh{\ii \tilde{\vec r} \vec k \left(1-\frac{\delta_\llg}{3}\right)}\\
=&\left(1+\frac{47}{21}\delta_\llg\right) P_\text{B}\left[k \left(1-\frac{\delta_\llg}{3}\right)\right]\\
\approx&\left(1+\frac{47}{21}\delta_\llg-\frac{\delta_\llg}{3}\frac{\derd \ln P}{\derd \ln k}\right) P_\text{B}(k)\\
\approx&\left(1+\frac{68}{21}\delta_\llg-\frac{\delta_\llg}{3}\frac{\derd \ln k^3 P}{\derd \ln k}\right) P_\text{B}(k)
\end{split}
\end{equation}
These results can also be found in \cite{Li:2014ss,Wagner:2015se,Wagner:2015th}, but we reproduce their derivation here for later reference.
The matter power spectra are shown in Fig.~\ref{fig:power}. In the left panel, we plot the power spectrum at the local wavenumber (circles) and interpolate to the global wavenumber (crosses). When taking ratios of the power spectra corresponding to the same global wavenumber, but ignoring the shift term, we see the $47/21$ enhancement, which is enhanced if we take ratios at the same comoving wavenumber. We see that there is a small amount of asymmetry between the overdense and underdense simulation, that originates is due to $\mathcal{O}\left(\deltal^2\right)$ effects which cancel once the symmetric derivative is taken. We show the symmetric derivatives on the right hand side of Fig.~\ref{fig:power}. Without the shift terms, the local power spectrum is enhanced by $47/21\deltal$ on large scales with further scale dependent enhancements on small scales arising from non-linearities. Once the shift term is included, the response is enhanced and becomes scale dependent as we show by the green triangles. We can model this logarithmic derivative using linear theory and find good agreement except for the BAO range, where the response is damped. Applying IR-resummation \cite{Senatore:2015ir, Baldauf:2015eq} to the linear power spectrum before taking the derivative reduces the deviations in the BAO regime. Note that this shift correction vanishes at the peak of the power spectrum $k\approx 0.015 \ihMpc$, where $\derd \ln P/\derd \ln k=0$.
%>>>>>>>>>>>>>>>>>>>>>>>>>>>>>>>>>>>>>>>>>>>>>>>>>>>>>>>>>>>>>>>>>>>>>>>>>>>>>>>>>>
%<<<<<<<<<<<<<<<<<<<<<<<<<<<<<<<<<<<<<<<<<<<<<<<<<<<<<<<<<<<<<<<<<<<<<<<<<<<<<<<<<<

%>>>>>>>>>>>>>>>>>>>>>>>>>>>>>>>>>>>>>>>>>>>>>>>>>>>>>>>>>>>>>>>>>>>>>>>>>>>>>>>>>>
%<<<<<<<<<<<<<<<<<<<<<<<<<<<<<<<<<<<<<<<<<<<<<<<<<<<<<<<<<<<<<<<<<<<<<<<<<<<<<<<<<<

\newpage
%===============================================================================
%===============================================================================
\section{Bias from the Curvature Simulations}\label{sec:bias}

In this Section we describe a simple way to measure the leading order bias. In general, galaxies at a given location are biased tracers of the dark matter density, velocity and gravitational field, evaluated in a spatial neighborhood of the point, along the whole past trajectory. The spatial dependence can be encoded in a series of progressively smaller spatial derivatives acting on the fields evaluated on the past trajectory of the galaxy. This leads to a general parametrization of the overdensity of halos (or galaxies) in terms of bias coefficients that take the following schematic form~\cite{McDonald:2009dh,Assassi:2014re,Senatore:2014eva,Mirbabayi:2014zca}
\be
\delta_\text{h}(x)\sim \sum_i b_i^\text{(E)} \,{\cal O}^{(i)}(x)+\epsilon^{(i)}\; ,
\ee
where $i$ is an index that runs over the various fields ($\delta,\partial_jv^i,\partial_i\partial_j\phi$ and products of them) and the order in perturbation theory at which these operators are evaluated. The coefficients $b_i^\text{(E)}$ are called Eulerian biases, while the $\epsilon^{(i)}$ are called stochastic Eulerian biases as they represent the part of the statistics of halos that is uncorrelated with the underlying dark matter fields. Recently operators involving baryonic fields \cite{Lewandowski:2015an,Angulo:2015eqa} and the effect of non-Gaussianities have been included as well~\cite{Angulo:2015eqa,Assassi:2015ef,Assassi:2015ga}.

The leading term in this bias expansion is the overdensity $\delta$, whose bias coefficient is normally denoted $b_1$. It is straightforward to understand how the number of galaxies depends on $\delta$. If we imagine that there is a spherical configuration of spatially constant overdensity at a given location, locally its effect can be encoded into a redefinition of the effective cosmology, where the Universe is a curved FLRW Universe with small scale density fluctuations (that ultimately collapse into galaxies). Therefore, we can write
\be\label{eq:main}
b_1^\text{(E)}=\frac{1}{\bar n}\frac{\partial n}{\partial \delta_\text{l}}\propto \frac{1}{\bar n}\frac{\partial n}{\partial \Omega_\text{K,L}}\;
\ee
where $\Omega_K$ is the curvature of the locally defined FLRW Universe. This naturally provides a way to measure the linear bias (or any bias that contributes for spherically symmetric configurations) using small-box curved-Universe simulations.  This was the line of reasoning that we developed in~\cite{Baldauf:2011ew} and implement here.

If this bias expansion is written down in Lagrangian space, then the bias parameters are denoted Lagrangian bias $b^\text{(L)}$, whereas if the same relation written down in the late time evolved density field, then the bias parameters are denoted Eulerian bias parameters $b^\text{(E)}$. The relation between the two can be deduced from the usual conversion between Lagrangian and Eulerian space
\be
\begin{split}
\Bigl[1+\delta(\vec x)\Bigr]\derd^3 x=&\derd^3 q\\
\Bigl[1+\delta_\text{h}^\text{(E)}(\vec x)\Bigr]\derd^3 x=&\Bigl[1+\delta_\text{h}^\text{(L)}(\vec q)\Bigr]\derd^3 q
\end{split}
\ee
yielding $b_1^\text{(E)}=1+b_1^\text{(L)}$ and $b_2^\text{(E)}=b_2^\text{(L)}+\frac{8}{21}b_1^\text{(L)}$~\cite{Mo:1996an}, where $b_2^\text{(E)}$ is the Eulerian bias coefficient associated with the square of the field $\delta^2$.

If one was able to describe galaxy formation from first principles, the number density of galaxies and the bias coefficients could be obtained analytically. Attempts in this direction have been formulated, using some simplifying assumptions. Although these theoretical mass functions make simplifying assumptions, they perform reasonably well for halos. We therefore use biases derived from these mass functions in order to have an estimate of the expected results. In the next subsection we are going to review the standard expressions for the bias coefficients in terms of a mass function. We stress that our method for obtaining bias from curved simulations is independent of the theoretical mass function: we simply compare the results obtained using Eq.~\eqref{eq:main} with the ones obtained using the standard power spectrum method -- namely cross correlating the dark matter density and the halo density fields. 
\subsection{Universal mass function}
 Let us start by considering the abundance of dark matter haloes.
The mass function quantifies the number density of collapsed cold dark matter haloes as a function of mass. 
This abundance is related to the distribution of collapsing overdense patches in the initial conditions via the Press-Schechter \cite{Press:1974fo} formalism. This formalism connects haloes to Lagrangian regions that exceed a critical collapse density $\delta_\text{c}$ and contain the same mass as the final halo under consideration. 
While there is a plentitude of theoretical mass functions with varying levels of theoretical footing and agreement with mass functions extracted from numerical simulations \cite{Press:1974fo,Sheth:1999mn,Tinker:2008to,Paranjape:2013ex}, whenever comparing to theoretical mass functions, we will employ the simple Sheth-Tormen mass function \cite{Sheth:1999mn} as a reference.
The halo mass function is related to the multiplicity
\be
\frac{\derd n}{\derd M}=\nu f(\nu)\frac{\bar \rho}{M^2}\frac{\derd \log \nu}{\derd \ln M}\; ,
\ee
where $\nu=\delta_\text{c}^2/\sigma_M^2$ and
\be
\nu f(\nu)=A(p)\left(1+\frac{1}{(q\nu)^p}\right)\sqrt{\frac{q\nu}{2\pi}}\eh{-\frac{q\nu}{2}}.
\label{eq:stmassfct}
\ee
The above functional form with fitting parameters $p=0.15$ and $q=0.73$ provides a reasonable fit to the abundance of haloes in our simulations, as can be seen in Fig.~\ref{fig:mfun}. The number density of haloes is not constant in space, but correlated with the background long wavelength density fluctuations. This can be easily understood from the point of view of the mass function from the fact, that a overdensity of a given radius is more likely to cross the density threshold if it is situated in a long wavelength overdensity, effectively reducing the threshold to $\delta_\text{c}-\deltal$.  Therefore, an expression for the bias parameter can be obtained from a theoretical bias function by changing $\delta_\text{c}\to \delta_\text{c}-\delta_\text{l}$.
As a reference for our measurements, we will consider the biases derived from the ST mass function.
The Lagrangian bias parameter can be derived from the universal mass function expressed in terms of the peak height
\be
\begin{split}
b_\delta^\text{(L)}=&\frac{1}{n}\frac{\partial n}{\partial \delta_\text{l}}=\frac{1}{n}\frac{\partial n}{\partial \nu}\frac{\partial \nu}{\partial \delta_\text{l}}
=-\frac{2\delta_\text{c}}{\sigma_M^2}\frac{1}{n}\frac{\partial n}{\partial \nu}=-\frac{2\nu}{\delta_\text{c}}\frac{1}{n}\frac{\partial n}{\partial \nu}\; ,\\
b_{2}^\text{(L)}=&\frac{1}{\bar{n}}\frac{\partial^2 n}{\partial \delta_\text{l}^2}=\frac{4}{\bar
n}
\frac{\nu^2}{\delta_c^2}\frac{\partial^2 n}{\partial \nu^2}+\frac{2}{\bar{n}}\frac{\nu}{
\delta_\text{c}^2}\frac{\partial n}{\partial \nu},
\end{split}
\ee
For the ST mass function introduced above, we have
\be
\begin{split}
\frac{1}{n}\frac{\partial n}{\partial \nu}=&-\frac{q\nu-1}{2\nu}-\frac{p}{\nu\left(1+(q\nu)^p\right)}\\
\frac{1}{n}\frac{\partial^2 n}{\partial
\nu^2}=&\frac{p^2+\nu p
q}{\nu^2\left(1+(q\nu)^p\right)}+\frac{(q\nu)^2-2q\nu-1}{4\nu^2}.
\end{split}
\ee

\subsection{Number Density and Bias from the Simulations\label{sec:number}}
\begin{figure}[tb]
\begin{center}
\includegraphics[width=0.49\textwidth]{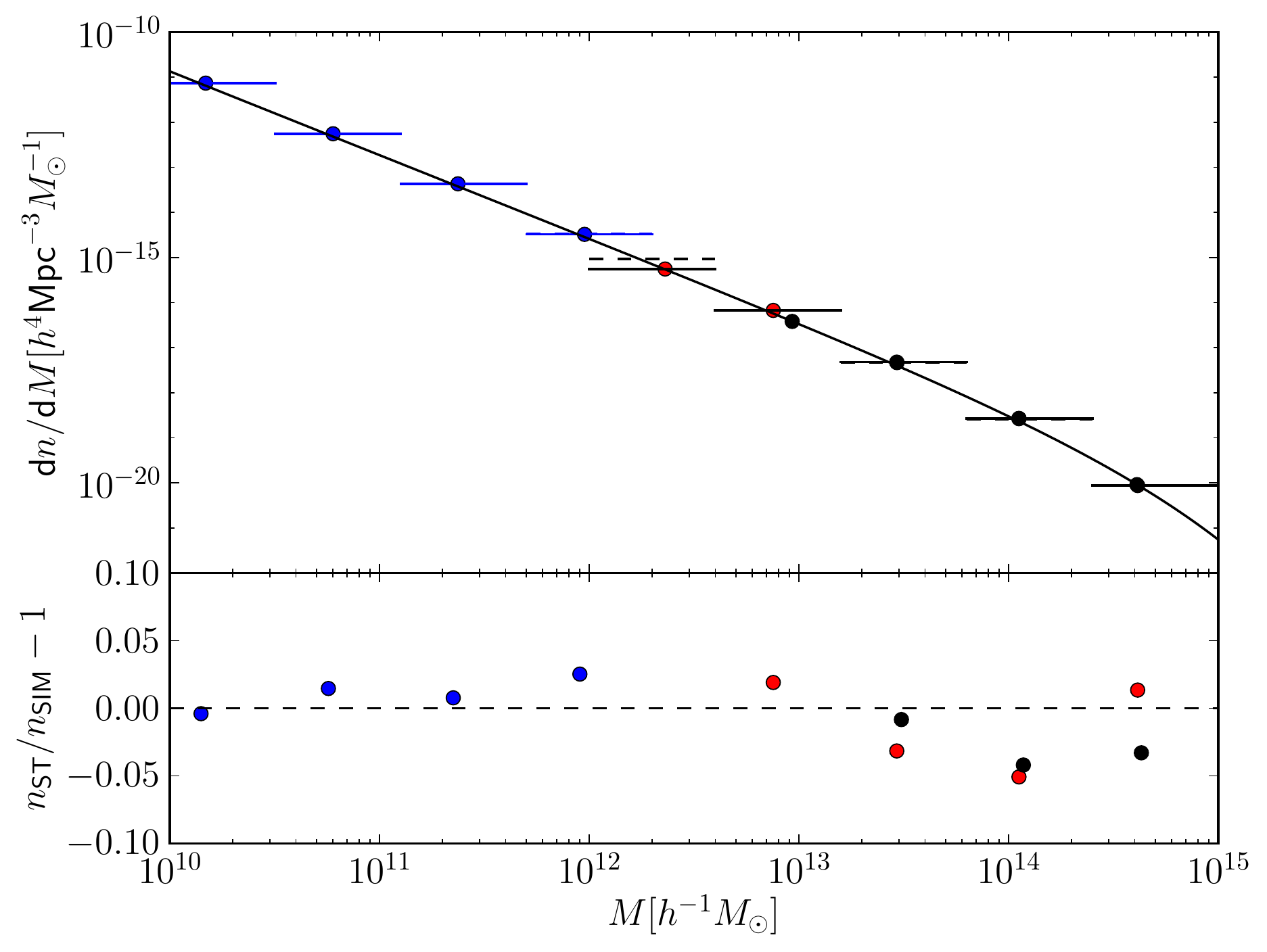}
\caption{Measured halo abundance in the S, M and L simulation. We overplot the ST mass function for $p=0.15$ $q=0.73$. The horizontal lines give the width of the mass bins. The ordinate of the dashed lines is the theoretical mass function integrated over the bin and the ordinate of the solid lines is the measured abundance. \emph{Lower panel: }Fractional deviation between the theoretical and measured mass function. Note that the deviation for the lowest mass bins from the M and L simulation exceed the scale of the lower panel.}
\label{fig:mfun}
\end{center}
\end{figure}

For the M and L boxes we are splitting the mass range between $10^{12} \hMs$ and $10^{15}\hMs$ into five logarithmic mass bins. There are no haloes for the lowest mass bin in the L simulation.
For the S simulation, we are splitting the mass range between $8\tim{9}\hMs$ and $2\tim{12}\hMs$ into four logarithmic mass bins.
The number density in the fiducial runs is shown in Fig.~\ref{fig:mfun}, where we also show a comparison to the theoretical ST mass function. The agreement is quite good, except for the lowest bin in the M simulation and for the lowest bin in the L simulation (which corresponds to the second bin of the M simulation), whose fractional deviation from the mass function exceeds the scale of the lower panel.
%>>>>>>>>>>>>>>>>>>>>>>>>>>>>>>>>>>>>>>>>>>>>>>>>>>>>>>>>>>>>>>>>>>>>>>>>>>>>>>>>>>
%<<<<<<<<<<<<<<<<<<<<<<<<<<<<<<<<<<<<<<<<<<<<<<<<<<<<<<<<<<<<<<<<<<<<<<<<<<<<<<<<<<

%>>>>>>>>>>>>>>>>>>>>>>>>>>>>>>>>>>>>>>>>>>>>>>>>>>>>>>>>>>>>>>>>>>>>>>>>>>>>>>>>>>
\begin{figure}[t]
\centering
\includegraphics[height=5.9cm]{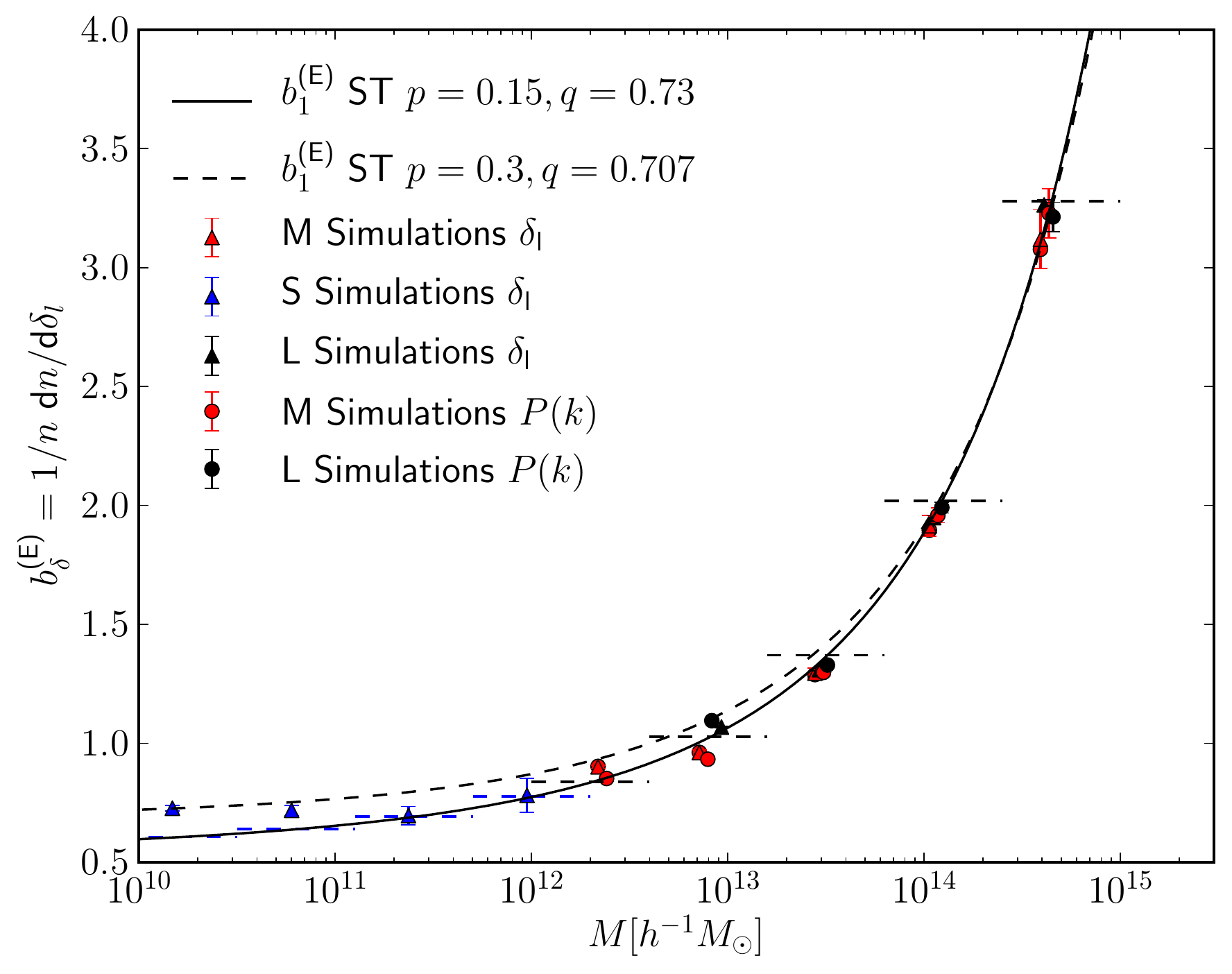}
\includegraphics[height=5.9cm]{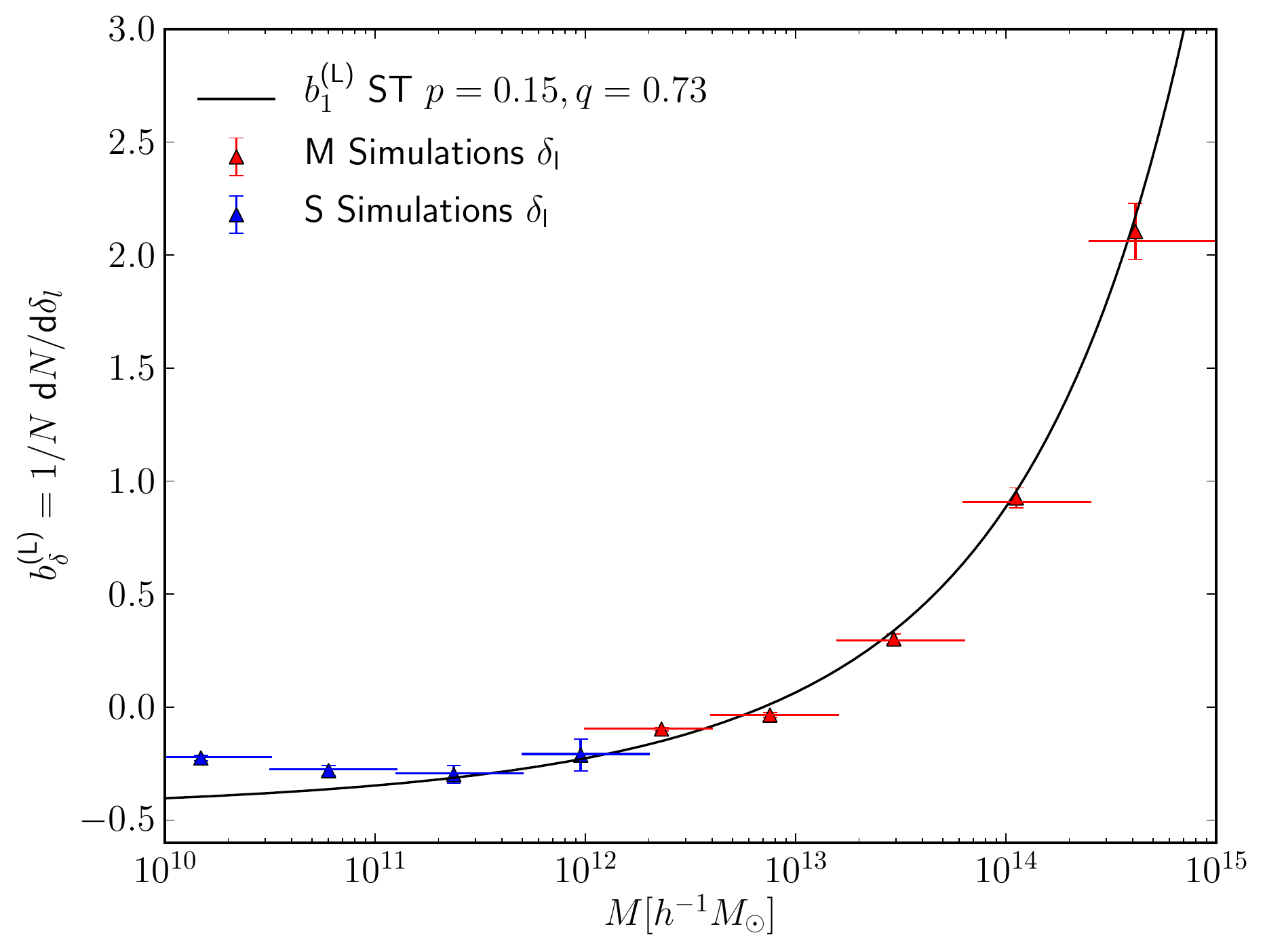}
\vspace{-0.4cm}
\caption{Bias parameters from the power spectrum and curvature method and the S (blue), M (red) and L (black) boxes. We are overplotting the bias function for two parameter choices in the ST mass function
\emph{Left panel: } Eulerian bias estimated according to Eq.~\eqref{eq:eulbias}. Horizontal dashed lines indicate the width of the mass bin and the value of the ($p=0.15$, $q=0.73$) mass function averaged over the bin. \emph{Right panel: }Lagrangian bias estimated according to Eq.~\eqref{eq:lagbias}. 
At the low mass end there is a clear deviation from the ST bias function, that was previously reported in \cite{Seljak:2004mn}. A ratio to the theoretical bias function is shown in Fig.~\ref{fig:compareest}.}
\label{fig:deltabias}
\vspace{-0.3cm}
\end{figure}
%<<<<<<<<<<<<<<<<<<<<<<<<<<<<<<<<<<<<<<<<<<<<<<<<<<<<<<<<<<<<<<<<<<<<<<<<<<<<<<<<<<

The linear bias is the response of the halo number density to the presence of a long wavelength mode. Thus it can be estimated from the number counts in over- and underdense simulations $N_\pm$ and the fiducial simulation as follows
\be
b^\text{(E)}_{\delta,\text{sim}}=\frac{N_{\delta_+}/V_{\delta_+}-N_{\delta_-}/V_{\delta_-}}{N_\text{f}/V_\text{f}(\delta_+-\delta_-)}.
\label{eq:eulbias}
\ee
Note that we have symmetrically spaced overdensities, such that $\delta_{\text{l},\pm}=\pm |\deltal|$.
Here we accounted for the fact that the local box size in global comoving coordinates is different from the fiducial comoving box size, i.e., $V_{\delta_\pm}=V_\text{f}(1-\delta_\pm)$.
%>>>>>>>>>>>>>>>>>>>>>>>>>>>>>>>>>>>>>>>>>>>>>>>>>>>>>>>>>>>>>>>>>>>>>>>>>>>>>>>
\begin{figure}
\centering
\includegraphics[width=0.49\textwidth]{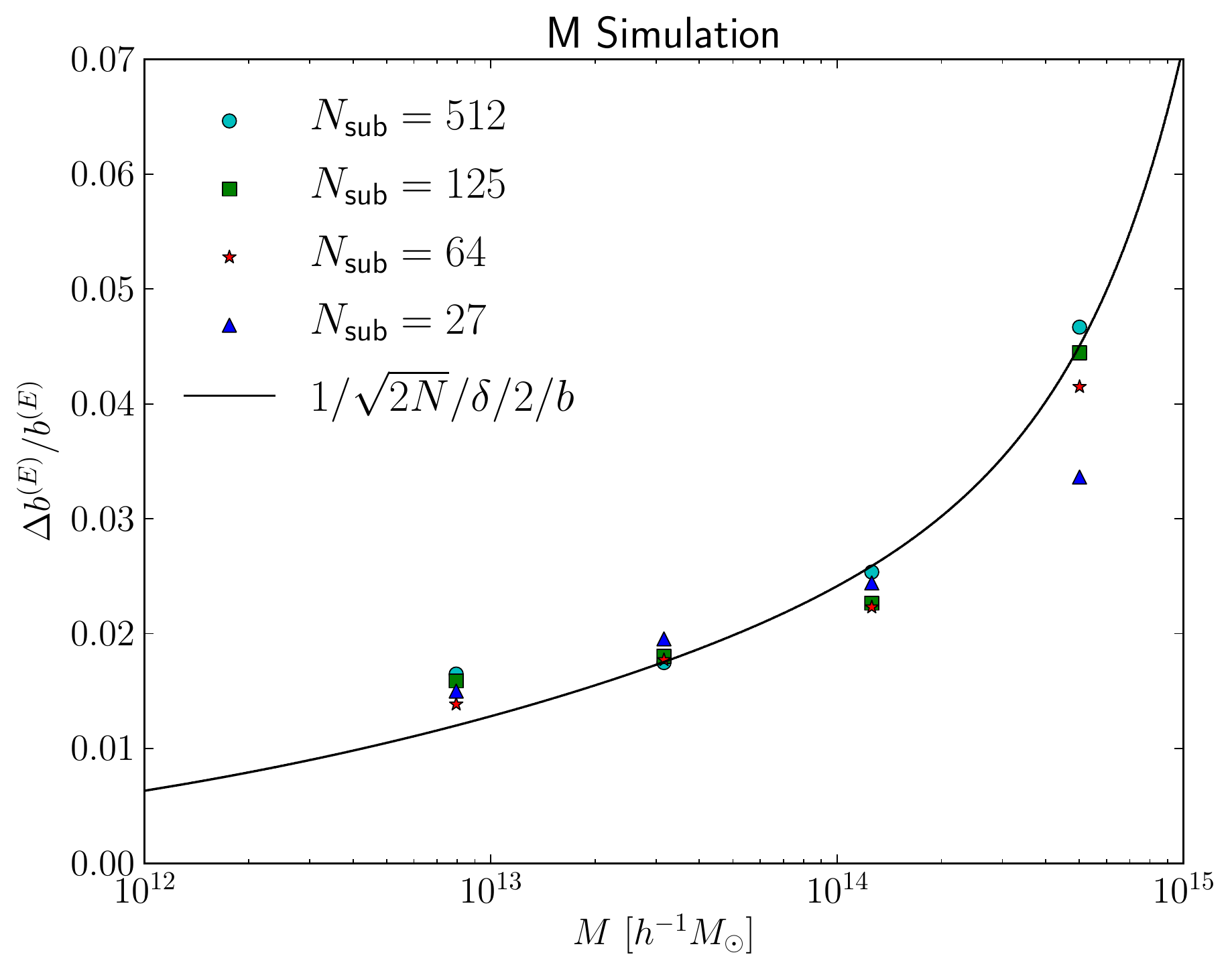}
\includegraphics[width=0.49\textwidth]{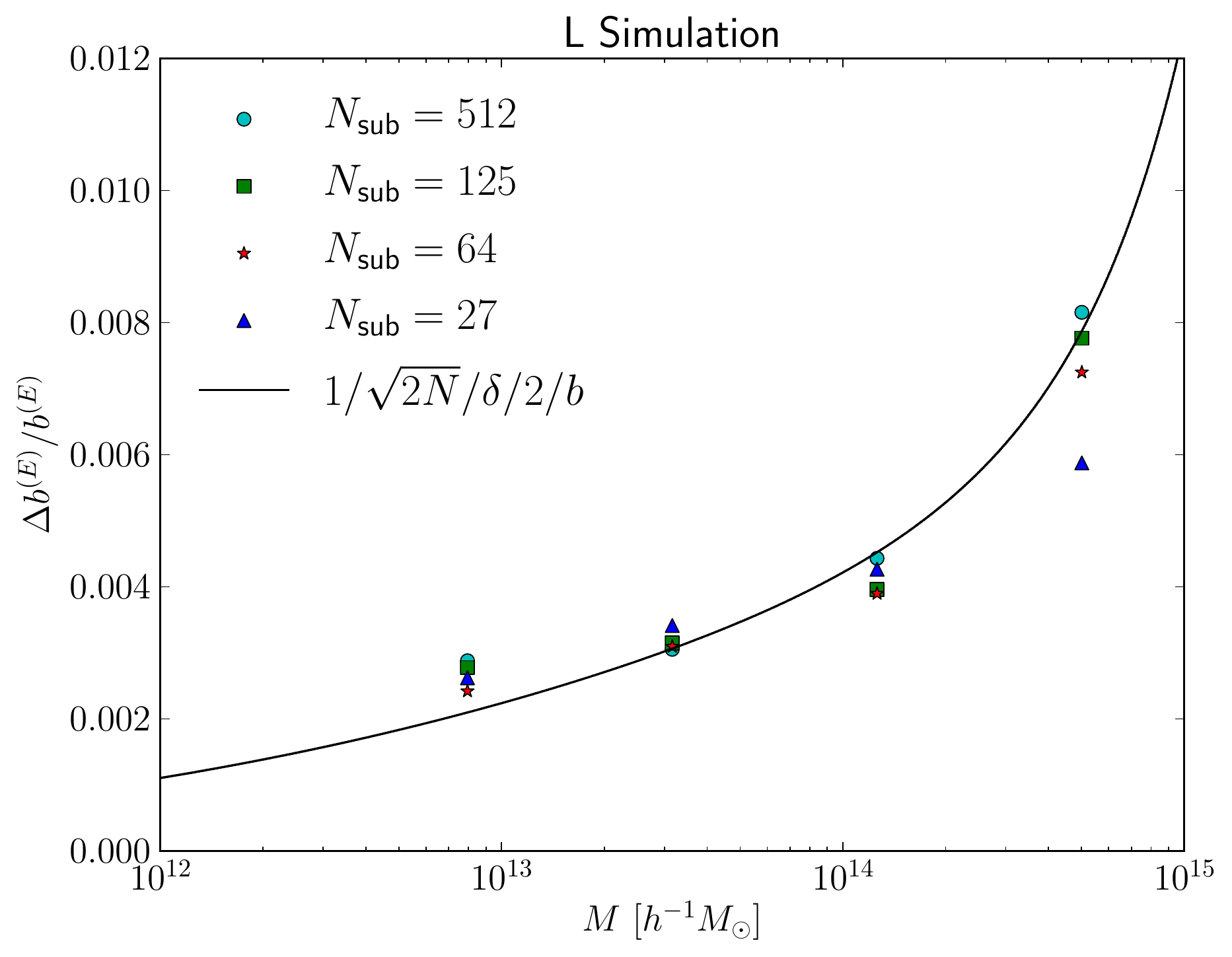}
\caption{Relative error on the bias parameters. For this purpose we split the L box into $N_\text{sub}$ subvolumes and calculate the response of the number of haloes to the long mode in each of the subvolumes. We then use jackknife resampling to estimate the variance of the bias for the full volume. We obtain stable results for the variance when changing the number of subvolumes. For the M box, the errors from the splitting of the L box were rescaled by the square root of the ratio of the box volumes. The black line is given by Eq.~\eqref{eq:deltalerror} divided by 4.}
\label{fig:relerr}
\end{figure}
%<<<<<<<<<<<<<<<<<<<<<<<<<<<<<<<<<<<<<<<<<<<<<<<<<<<<<<<<<<<<<<<<<<<<<<<<<<<<<<<<<<
The variance of the bias parameter can be roughly estimated from the Poisson error on the halo counts
\be
\begin{split}\label{eq:deltalerror}
\left(\Delta
b^\text{(E)}\right)^2=&\frac{1}{4\deltal^2}\left[
\left(
\frac{V_\text{f}}{V_{\delta_+}}\frac{1}{N_\text{f}}\right)^2\left(\Delta{N_{\delta_+}}\right)^2
+\left(\frac{V_\text{f}}{V_{\delta_-}}\frac{1}{N_\text{f}}\right)^2\left(\Delta{N_{\delta_-}}\right)^2
+\left(\frac{N_{\delta_+} \frac{V_\text{f}}{V_{\delta_+}}-N_{\delta_-}\frac{V_\text{f}}{V_{\delta_-}}}{
N_\text{f}^2}\right)^2\left(\Delta{N_\text{f}}\right)^2
\right]\\
\approx&\frac{1}{2 N_\text{f}\deltal^2}
\end{split}
\ee
This error estimate assumes that the number of haloes in the overdense, fiducial and underdense simulations is not correlated, which is clearly overly simplistic and thus certainly overestimates the true error. 
For this reason we split the L box into $N_\text{sub}$ subvolumes and count the number of haloes in each subvolume, calculate the bias according to Eq.~\eqref{eq:eulbias} for each of the subvolumes and perform a jackknife resampling to estimate the variance of the estimator.
In these jackknife resampling we find that the functional form of the error estimate seems to work but the amplitude is a factor of 4 too large (see Fig.~\ref{fig:relerr}).
Alternatively, we could have also considered the response of the number, rather than the number density
\be
b^\text{(L)}_{\delta,\text{sim}}=\frac{N_{\delta_+}-N_{\delta_-}}{N_\text{f}(\delta_+-\delta_-)},
\label{eq:lagbias}
\ee
which is thus a measure of the Lagrangian bias.\\
In Fig.~\ref{fig:deltabias} we show the Lagrangian and Eulerian bias measured from the numerical derivative between the flat and curved simulation boxes. At the high mass end the Sheth-Tormen bias function is a good description of the data while it underpredicts the bias of haloes with masses below the characteristic mass $M_* \approx 5\tim{12}\hMs$ . We will discuss the agreement of these measurements with the standard power spectrum method in much more detail below in Sec.~\ref{sec:powspecmethod}.
\subsection{Variance bias}
\begin{figure}
\includegraphics[height=5.9cm]{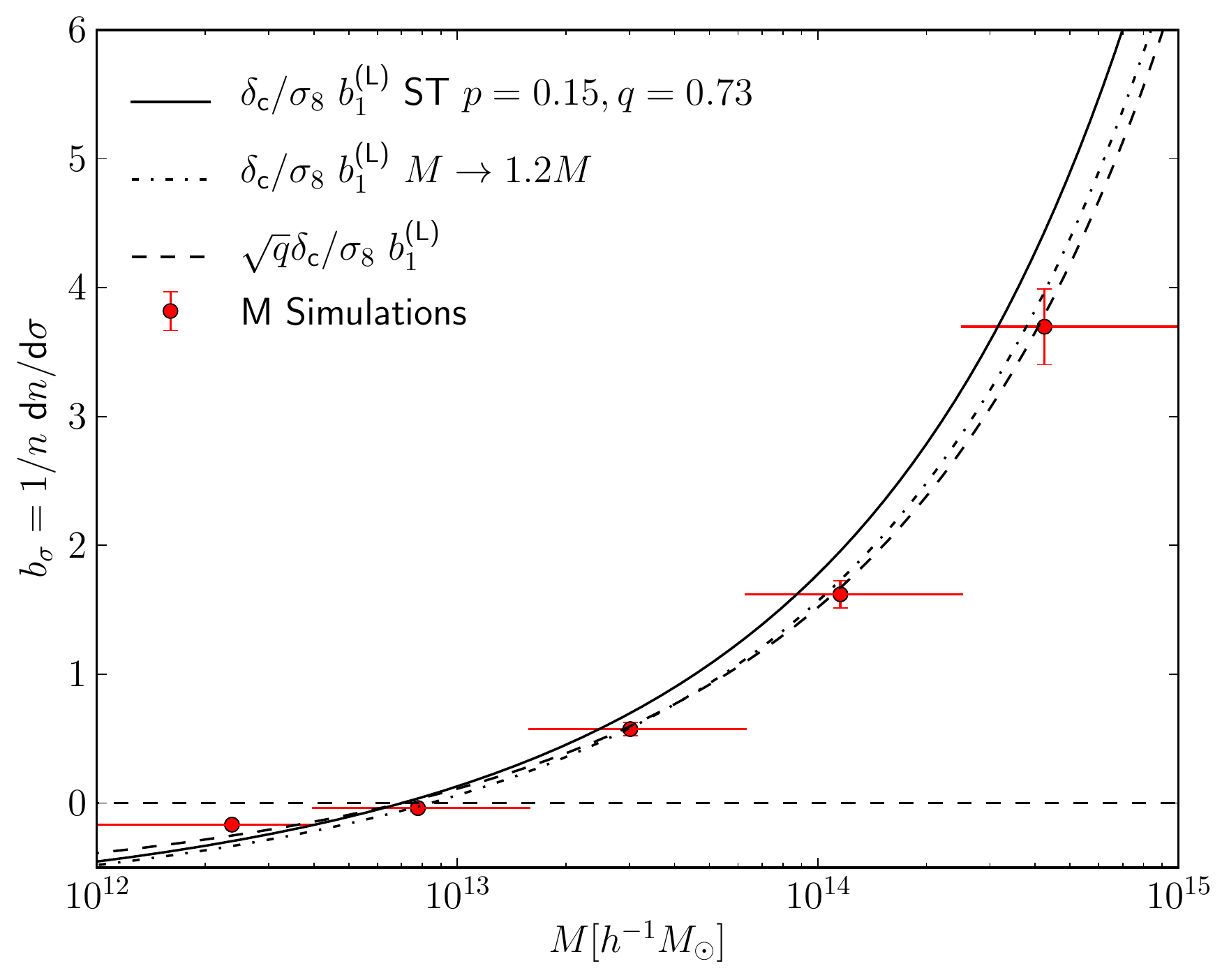}
\includegraphics[height=5.9cm]{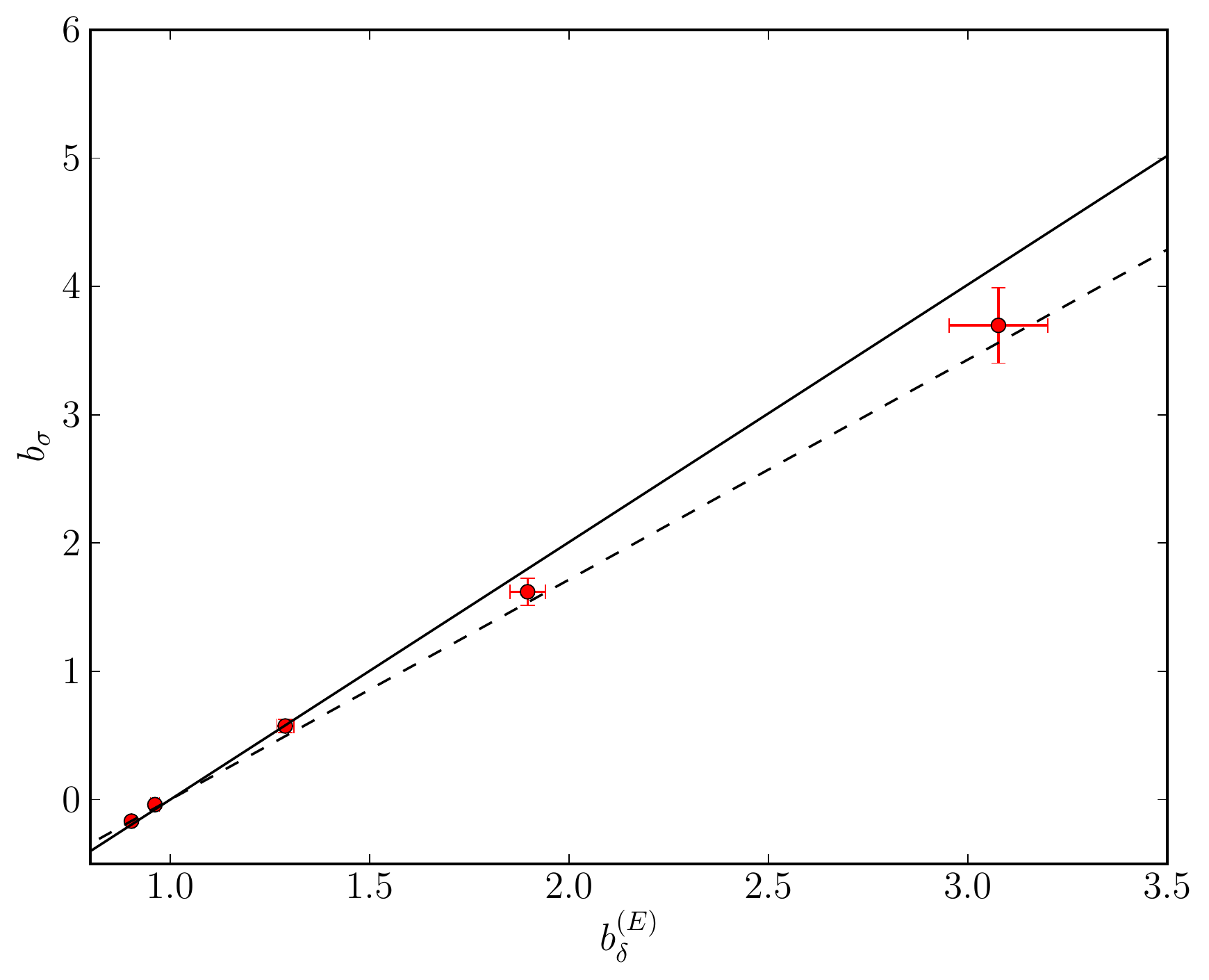}
\caption{Response of halo number counts to a change in the variance. 
We are overplotting the na\"\i ve scaling Eq.~\eqref{eq:bsigbdel} as well as versions that are obtained by accounting for the spherical collapse factor $q$ or rescaling the mass in the prediction.
\emph{Left panel: }$b_\sigma$ as a function of mass, where the horizontal lines indicate the width of the bin. \emph{Right panel: } $b_\delta^\text{(E)}-b_\sigma$ relation. We see significant evidence for the slope being overestimated by the derivation based on a universal mass function but find good agreement if $\delta_\text{c}$ in Eq.~\eqref{eq:bsig8} is replaced by~$\sqrt{q}\delta_\text{c}$.}
\label{fig:variancebias}
\end{figure}
Let us now consider the response of halo number density to a change in the background variance, which is related to the bias induced by primordial non-Gaussianities of the local form~\cite{Baldauf:2011ew}
\be
b_\sigma=\frac{1}{n}\frac{\derd n}{\derd \sigma}=\frac{N_{\sigma_+}-N_{\sigma_-}}{N_\text{f}(\sigma_+-\sigma_-)}.
\ee
Here there is no distinction between the Eulerian and Lagrangian bias at first order.
In the case of a universal mass function, defining $\sigma_M=\gamma_M\sigma_8$, the linear response of the tracer density with respect to a change in the normalization $\sigma_8$ yields
\be
b_{\sigma_8}=\frac{1}{n}\frac{\derd n}{\derd \sigma_8}=\gamma_M\frac{1}{n}\frac{\derd n}{\derd \nu}\frac{\derd \nu}{\derd\sigma_M}
=-\frac{\sigma_M}{\sigma_8}\frac{2\delta_\text{c}^2}{\sigma_M^3}\frac{1}{n}\frac{\derd n}{\derd \nu}=-\frac{2\nu}{\sigma_8}\frac{1}{n}\frac{\derd n}{\derd \nu}
\label{eq:bsigbdel}
\ee
Thus, for a universal mass function,
\be
b_{\sigma_8}=\frac{\delta_\text{c}}{\sigma_8}b_\delta^\text{(L)}.
\label{eq:bsig8}
\ee
As we show in Fig.~\ref{fig:variancebias}, this overpredicts the variance bias especially for large halo masses. This motivated the $q$-model \cite{Grossi:2009mn}, in which one replaces $\delta_\text{c}\to \sqrt{q}\delta_\text{c}$, where $q$ is the same parameter appearing in the ST mass function Eq.~\eqref{eq:stmassfct}. In the framework of the ST mass function Eq.~\eqref{eq:stmassfct} this modification makes sense, since all the $\nu \propto \delta_\text{c}^2$ occurring in this framework are rescaled by $\nu \to q \nu$, which is thus equivalent to rescaling $\delta_\text{c}\to \sqrt{q}\delta_\text{c}$.
It was furthermore argued in the literature \cite{Desjacques:2010pr} that this effect goes away if one uses Spherical-Overdenisty (SO) instead of FoF haloes. The latter effect can be emulated by reducing the halo mass by $20 \%$ or increasing the ordinate value of the theoretical prediction by the same amount. 
\par
The above relations can be connected to the non-Gaussian bias \cite{Dalal:2008hh}. In local quadratic non-Gaussianity, the presence of quadratic interactions in the inflationary action leads to a coupling between the short wavelength density variance and the long wavelength Gaussian gravitational potential~$\varphi$ in the initial conditions \cite{Dalal:2008hh}
\be
 \Phi_\text{nG}=\varphi+f_\text{NL} \varphi^2 \ \ \Rightarrow \ \ \sigma_{M,\text{nG}}=\sigma_{M,\text{G}}(1+2 f_\text{NL}\varphi)\; .
\ee
This modulation of the variance leads to a modulation of halo formation by the long wavelength potential, which has motivated a multivariate bias expansion \cite{Giannantonio:2010}
\be
\delta_\text{h}=b_\delta^\text{(E)}\delta+b_\varphi \varphi\; .
\ee
From the dependence of the variance on the Gaussian potential ${\derd \sigma_M}/{\derd \varphi}=2 f_\text{NL} \sigma_M$, and assuming a universal mass function, one can then deduce the non-Gaussian bias parameter
\be
b_\varphi=2 f_\text{NL} \sigma_M b_{\sigma_M}=2f_\text{NL}\delta_\text{c} b_\delta^\text{(L)}=2f_\text{NL}\delta_\text{c} \left(b_\delta^\text{(E)}-1\right).
\label{fig:phibias}
\ee
Based on our response measurements the slope in the $b_\varphi-b_\delta^\text{(L)}$ relation is overestimated by $q^{-1/2}\approx 1.17$ for $q=0.73$ and $q^{-1/2}\approx 1.19$ for $q=0.707$. This means in turn, that $f_\text{NL}$ estimated using this relation is actually underestimated. 
The linear scale dependent bias in the presence of local non-Gaussianity has to be equal to the derivative of the mass function with respect to the potential, thus the measurements presented here should correspond to the true coefficients measured in non-Gaussian simulations. 
One can apply the corresponding corrections to existing measurements. For instance, using FoF haloes and estimating the amplitude of local $f_\text{NL}$ using the relation Eq.~\eqref{fig:phibias},~\cite{Hamaus:2011op} find a $f_\text{NL}$ that is $20\%$ lower than the value that was used to run the simulation. As we argued, this deviation can be explained by the reduced response of the mass function to the amplitude of fluctuations.

%===============================================================================
\subsection{Comparison to the Standard Power Spectrum Method}\label{sec:powspecmethod}
%>>>>>>>>>>>>>>>>>>>>>>>>>>>>>>>>>>>>>>>>>>>>>>>>>>>>>>>>>>>>>>>>>>>>>>>>>>>>>>>
\begin{figure}[t]
\centering
\includegraphics[width=0.49\textwidth]{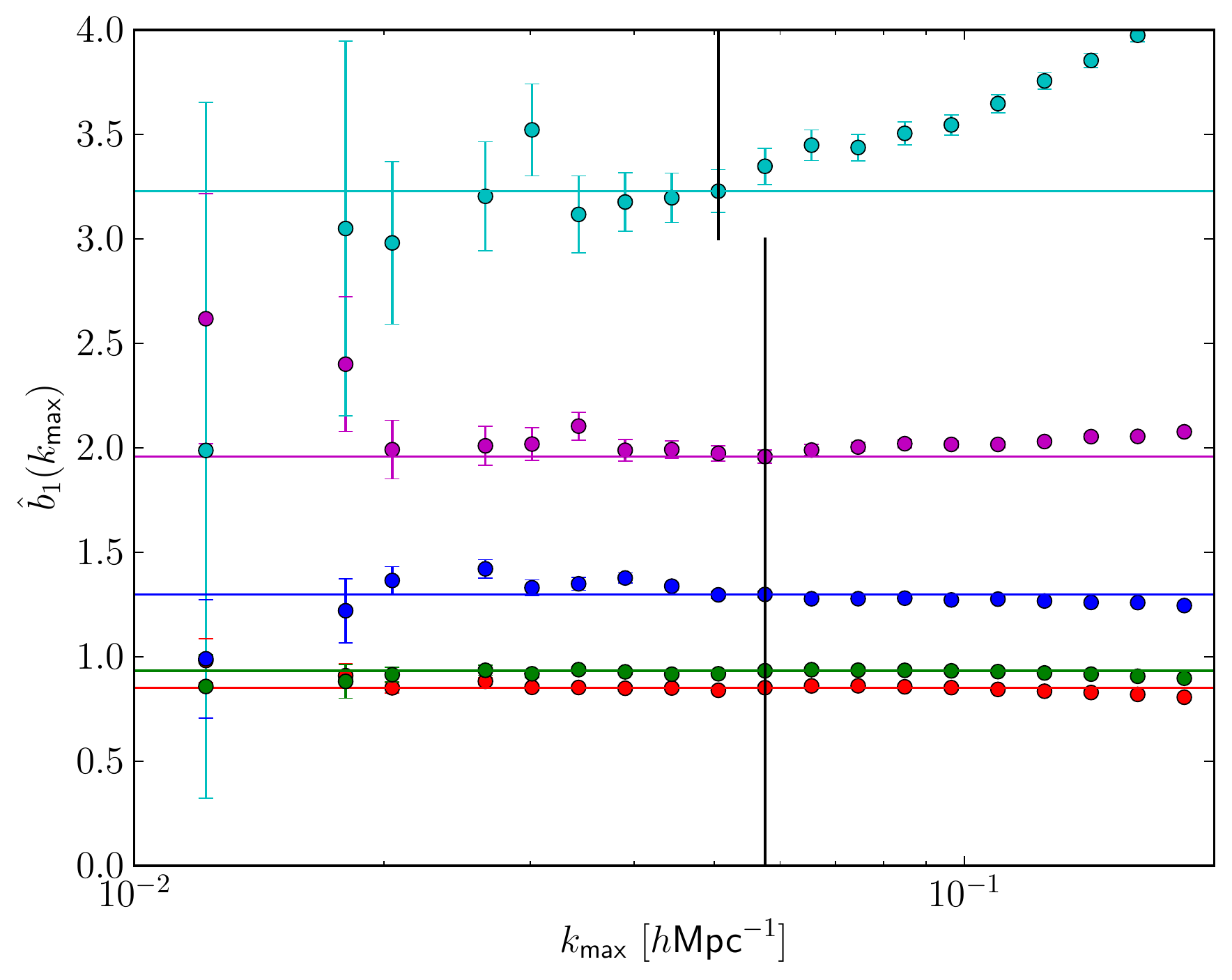}
\includegraphics[width=0.49\textwidth]{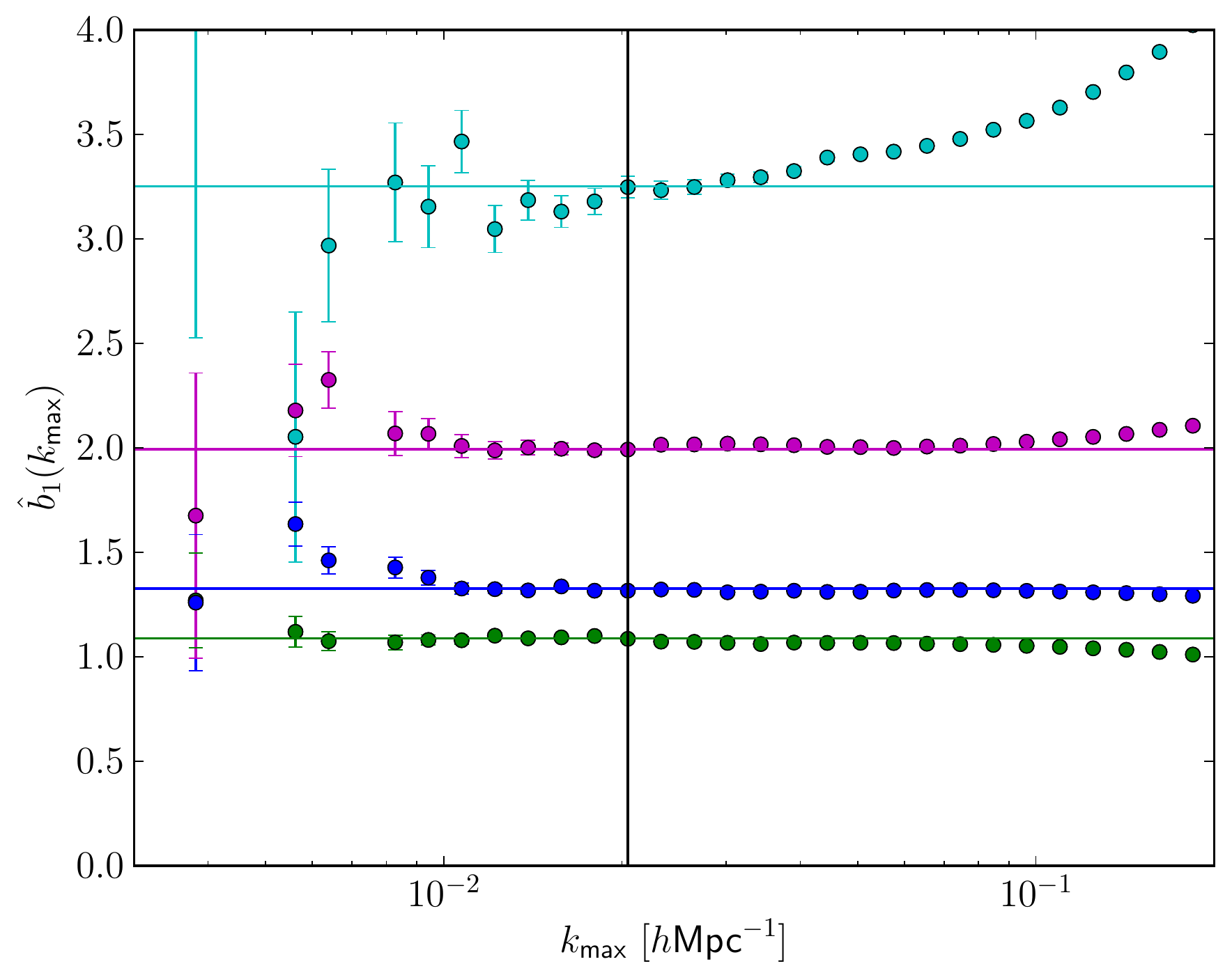}
\vspace{-0.4cm}
\caption{Cumulative constraints on the bias from the power spectrum method according to Eq.~\eqref{eq:b1powerest}. The horizontal lines show the resulting constraint and the vertical lines show the maximum wavenumber considered for the constraint. Mass increases from bottom to top. \emph{Left panel: }M simulation $k_\text{max}=0.06 \ihMpc$ (except for the highest mass bin, for which it is $k_\text{max}=0.05 \ihMpc$ since there is a clear step beyond this wavenumber) \emph{Right panel: }L simulation $k_\text{max}=0.018 \ihMpc$ }
\label{fig:b1powerconst}
\vspace{-0.3cm}
\end{figure}
%<<<<<<<<<<<<<<<<<<<<<<<<<<<<<<<<<<<<<<<<<<<<<<<<<<<<<<<<<<<<<<<<<<<<<<<<<<<<<<<
We compare the results of the curvature procedure described above to the
standard procedure of measuring the first order bias parameter from the power spectrum.
In order to avoid the spurious shotnoise contamination in the halo-halo power
spectrum $P_\text{hh}(k)$, we will only consider the cross power spectrum
between haloes and matter $P_\text{hm}(k)$ and define the scale dependent estimated halo
bias as\footnote{We stress that the bias is scale {\it independent}: there are many scale-independent bias parameters, each one corresponding to higher derivatives of the fields. The notation of Eq.~\eqref{eq:scale-dep} should be understood as the result of the estimation of the scale-independent linear bias using modes of wavenumber $k$.}
\be\label{eq:scale-dep}
b(k)=\frac{P_\text{mh}(k)}{P_\text{mm}(k)}\ .
\ee
The linear bias is then estimated upon minimizing
\be
\chi^2=\sum_{k=0}^{k_\text{max}} \frac{\bigl(b(k)-b_1\bigr)^2}{\Delta^2 b(k)}
\ee
yielding the following estimator
\be
\hat b_{1,P}=\sum_{k=0}^{k_\text{max}} \frac{b(k)}{\Delta^2 b(k)}/\sum_{k=0}^{k_\text{max}} \frac{1}{\Delta^2 b(k)}
\label{eq:b1powerest}
\ee
The variance of $b(k)$ is given by
\be
\left(\Delta b(k)\right)^2=\frac{1}{N_k} \frac{1}{\bar n P_\text{lin}(k)}.
\ee
Comparing to the variance of the cross-power spectrum
\be
\left(\Delta P_\text{hm}\right)^2(k)=\frac{1}{N_k}\left[ P_\text{lin}(k)\left(b_1^2 P_\text{lin}(k)+\frac{1}{\bar n}\right)+b_1^2 P^2_\text{lin}(k)\right]
\ee
we clearly see that the ratio removes the dominant source of error on large scales, the cosmic variance. We have compared the above error estimates to the variance of the power spectra measured in a suite of 16 simulations and found very good agreement.
The resulting error on the constraint is then given by
\be
\Delta \hat b_{1,P}=\left(\sum_{k=0}^{k_\text{max}}\frac{1}{(\Delta b(k))^2}\right)^{-1/2}=\left(\bar n\sum_{k=0}^{k_\text{max}}N_k P(k)\right)^{-1/2}
\ee
For a finite volume with a long mode given by $\delta_\text{l}(\vec q)=V \deltal \delta^\text{(K)}_{\vec k,0}$ we have $P=1/V \la\delta(\vec k)\delta(-\vec k)\ra=V \deltal^2$. This yields for the error of the bias measured using a single long mode (for instance the constant box mode)
\be\label{eq:singlemodeerror}
\left(\Delta \hat b_{1,P}\right)^2=\frac{1}{N_\text{h}\delta^2_\text{l}}\; .
\ee
%>>>>>>>>>>>>>>>>>>>>>>>>>>>>>>>>>>>>>>>>>>>>>>>>>>>>>>>>>>>>>>>>>>>>>>>>>>>>>>>
\begin{figure}[t]
\begin{center}
\includegraphics[width=0.59\textwidth]{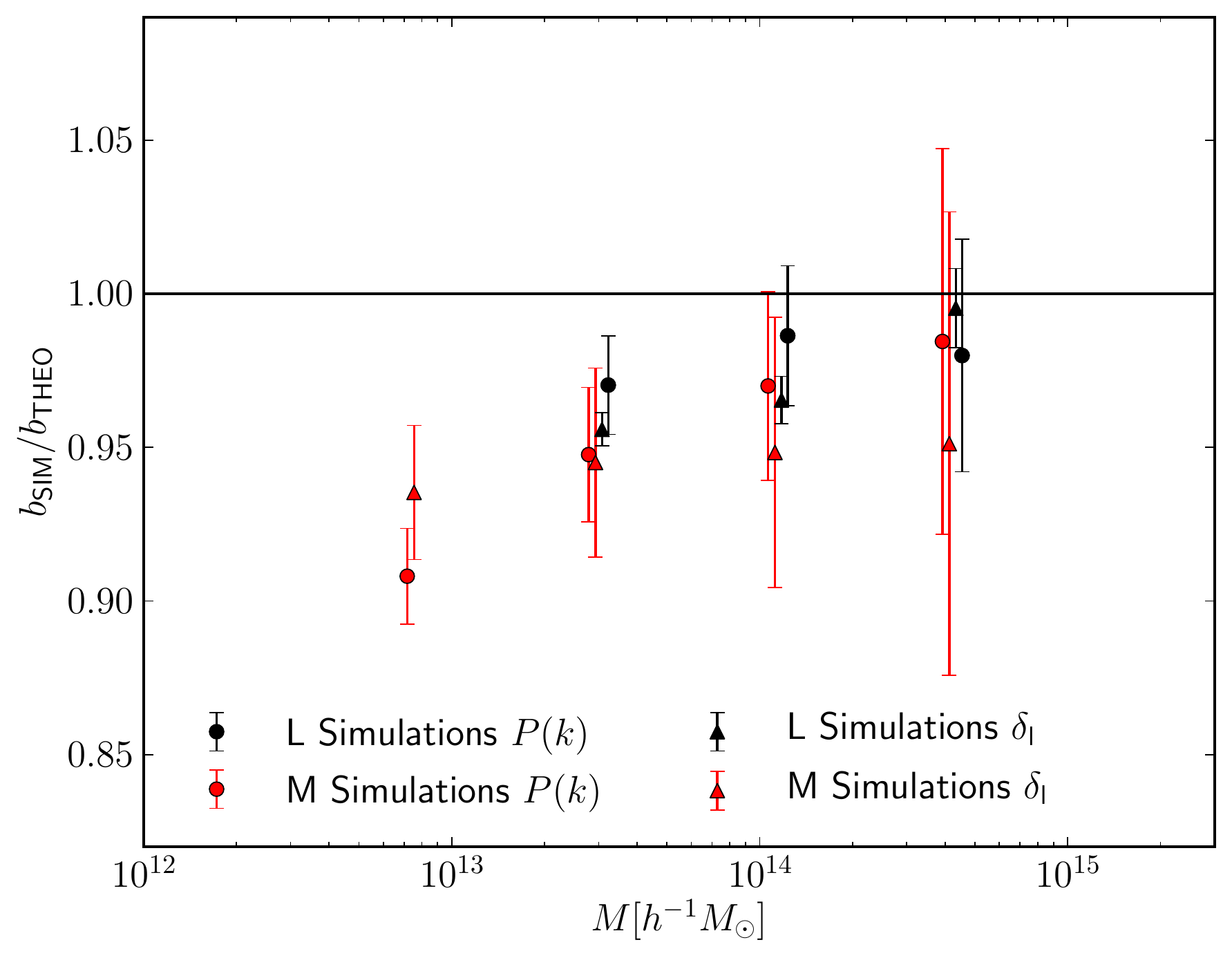}
\vspace{-0.4cm}
\caption{Fractional deviation between the measured bias parameters and the ST ($p=0.15,\ q=0.73$) bias prediction for the corresponding mass bin. Results from M simulations are colored in red and results from the L simulations in black. The circles show the constraints obtained using the power spectrum method and the triangles constraints obtained using the curvature method. The biases obtained from the two methods are generally in agreement but there is some mild tension for the L simulation, we argue about reasons for this discrepancy in the main text. We do not show the respective lowest bins of the simulations since they deviate from the mean trend in this plot and are likely plagued by incompleteness. We show two sigma errors. Note that the point of this plot is the mutual agreement between the two methods, not the agreement with the reference bias function. 
}
\label{fig:compareest}
\end{center}
\end{figure}
%<<<<<<<<<<<<<<<<<<<<<<<<<<<<<<<<<<<<<<<<<<<<<<<<<<<<<<<<<<<<<<<<<<<<<<<<<<<<<<<
This agrees with the estimate for the error of the bias inferred from the curvature method in Eq.~\eqref{eq:deltalerror} up to a factor of 2, which indeed is what expected because in order to reproduce our curvature method with the power spectrum estimator using only fundamental modes, we would need two fundamental modes, one representing $\delta_{\text{l},+}$ and the other $\delta_{\text{l},-}$.\footnote{One could also use this argument  to calculate the $k_{\rm max}$ in the power spectrum method for which the two methods give the same constraint. For $\deltal=0.1$ as used here this is $k_{\rm max}\approx 0.06 \ihMpc$, i.e. what we used for the M box. For the L box with $k_{\rm max}=0.02 \ihMpc$, this estimate gives four times larger errors for the power spectrum method.}\\
The $b_1$ estimator in Eq.~\eqref{eq:b1powerest} and its errors are plotted in Fig.~\ref{fig:b1powerconst} as a function of the maximum wavenumber. Note the strong scale dependence of the estimator for the highest mass bin. This scale dependence is caused by non-linearities in the bias and requires one to carefully
limit the scales considered for the determination of the linear bias parameter.
We determine the linear bias using only the largest scales in the M box ($k_\text{max}=0.05 \ihMpc$) and L box ($k_\text{max}=0.018 \ihMpc$). These cutoff wavenumbers are shown by the
vertical lines in Fig.~\ref{fig:b1powerconst}.  As obvious in the left panel of Fig.~\ref{fig:b1powerconst} the small number of large scale modes in a small box increases the variance of the estimator and thus hampers a clean extraction of the linear bias parameters. Thus we do not even attempt to measure the bias parameters using the power spectrum from the S box.\\
The left panel of Fig.~\ref{fig:deltabias} compares the constraints of the
standard method (green points with error bars) to the results of the curvature
method and Fig.~\ref{fig:compareest} shows their relative deviation from the theoretical bias function.
There is clearly some mild tension between the two approaches in Fig.~\ref{fig:deltabias}.
This has various reasons. For the bias estimate from the power spectrum we are neglecting higher order local bias terms as well as $k^2$ terms that would be predicted in the peak model or more generally in effective field theory treatments of bias. From the scale dependence measured in a larger suite of simulations \cite{Saito:2014un}, we infer that the scale dependence at scales of $k\approx 0.02 \ihMpc$ is below $1 \%$ and at scales of $k\approx 0.05 \ihMpc$ is below $2\%$. In the mass range relevant for the M and L boxes, the corrections are typically positive and would thus reduce the tension.
It is a bit harder to assess the systematic effects on the curvature method, since our mapping is only correct to first order. We are performing a second order accurate derivative and there might thus be spurious effects at third order. The local bias model predicts $b_3$ to be positive for $M<10^{13}\hMs$, becomes negative and crosses zero again at $M\approx 10^{15}\hMs$. It would thus likely reduce biases estimated from the curvature method at the order considered here, and correcting for this effect should reduce the tension.\footnote{Though here we are not interested here in investigating the performance of the ST mass function to extract the bias coefficients, tests of the accuracy of peak background split, i.e. the correspondence between the derivative of the number density of haloes with respect to a long mode and the bias in the power spectrum, have so far relied on the assumption of a universal mass function.
Based on accurate fitting of the mass function and the resulting disagreement with the bias parameters inferred from the power spectrum, \cite{Manera:2010la} conclude that the peak background split fails at the $10 \%$ level. We do not find evidence for a breakdown at that level, and interpret their disagreement as probably due to a breakdown of the universality of the mass function.}
\par
Let us finish this section by considering the bias coefficients measured from the power spectrum method and their correspondence to the curvature method in more detail.
Implementing the third order local Eulerian bias model \cite{Fry:1992vr}~\footnote{The local Eulerian bias model is by now well understood not be a complete description of the clustering of tracers. A generalized approach was developed in~\cite{McDonald:2009dh}, and a complete EFT description has been provided in~\cite{Senatore:2014eva,Mirbabayi:2014zca}. However, for the illustrative point we try to make in this part of the section, restricting ourselves to a local Eulerian bias model is enough.}
\be
\delta_\text{h}(\vec x)=b_1^\text{(E)} \delta(\vec x)+\frac{1}{2!}b_2^\text{(E)}\left[\delta^2(\vec x)-\la\delta^2\ra\right]+\frac{1}{3!}b_3^\text{(E)}\delta^3(\vec x)
\ee
we have for the halo-matter cross power spectrum at one loop level
\be
P_\text{hm}(k)=b_1^\text{(E)} P_\text{1loop}(k)+\left(\frac{34}{21}b_2^\text{(E)}\sigma^2+\frac{1}{2}b_3\sigma^2\right)P(k)+b_2^\text{(E)} I_{12}(k)\; ,
\ee
where 
\be
I_{12}(k)=\int \frac{\derd^3 q}{(2\pi)^3}F_2(\vec q,\vec k-\vec q)P(q)P(|\vec k-\vec q|)\; .
\ee
On large scales where $P_\text{1loop}(k)\approx P(k)$ and $I_{12}(k)/P(k)\to 0$ we have
\be
\frac{P_\text{hm}(k)}{P_\text{mm}(k)}=b_1^\text{(E)}+\frac{34}{21}b_2^\text{(E)}\sigma^2+\frac{1}{2}b_3^\text{(E)}\sigma^2.
\label{eq:clustbias}
\ee
This would suggest that the bias measured from the power spectrum deviates from the linear bias parameter $b_1^\text{(E)}$ \cite{McDonald:2006cl}.
As was shown in \cite{Schmidt:2013pe} this is exactly what is measured by the peak-background split.
This can be seen by writing the variation of the number density of haloes due to modes up to third order
\be
\begin{split}
n=\bar n\biggl(1+b_1^\text{(E)} &\left(\delta^{(1)}+\delta^{(2)}+\delta^{(3)}\right)+\frac12 b_2^\text{(E)}[\delta^{(1)}]^2-\frac12 b_2^\text{(E)}\langle [\delta^{(1)}]^2\rangle\\
&+b_2^\text{(E)}\delta^{(1)}\delta^{(2)}-b_2^\text{(E)}\langle \delta^{(1)}\delta^{(2)}\rangle+\frac{1}{3!} b_3^\text{(E)}[\delta^{(1)}]^3\biggr)\; .
\end{split}
\ee
Here the mean density $\bar n$ and the average $\la \delta^2\ra$ are to be understood at a global level, i.e. in absence of the long mode.
The linear modes entering this expression can be split into a long and a short wavelength component $\delta^{(1)}=\delta^{(1)}_\text{s}+\delta^{(1)}_\text{l}$. If we are interested in the number density at the scale of the long mode, we have averaging over the short modes and keeping only linear terms in the long mode
\be
n_\text{l}=\bar n\left(1+b_1^\text{(E)} \deltal+\frac{34}{21}b_2^\text{(E)}\deltal\sigma_\text{s}^2+\frac{1}{2} b_3^\text{(E)}\sigma_\text{s}^2 \deltal \right)\; ,
\ee
where $\sigma_\text{s}^2=\langle \delta_\text{s}^2\rangle$.
Thus we have for the derivative with respect to the long mode
\be
\frac{1}{\bar n}\frac{\partial n_\text{l}}{\partial \deltal}=b_1^\text{(E)} +\frac{34}{21}b_2^\text{(E)}\sigma_\text{s}^2+\frac{1}{2} b_3^\text{(E)}\sigma_\text{s}^2\; ,
\ee
which agrees exactly with the linear bias coefficient in the clustering in Eq.~\eqref{eq:clustbias}. We verify that both the power spectrum and the curvature method measure the same quantity. However, as usual, when these measurements are interpreted as in terms of bias coefficients, one needs to carefully account for the effect of the higher order biases at a given order in perturbation theory (i.e. the so called renormalization).

\subsection{Second order bias from clustering and Super sample variance of halo statistics}
\begin{figure}
\includegraphics[width=0.49\textwidth]{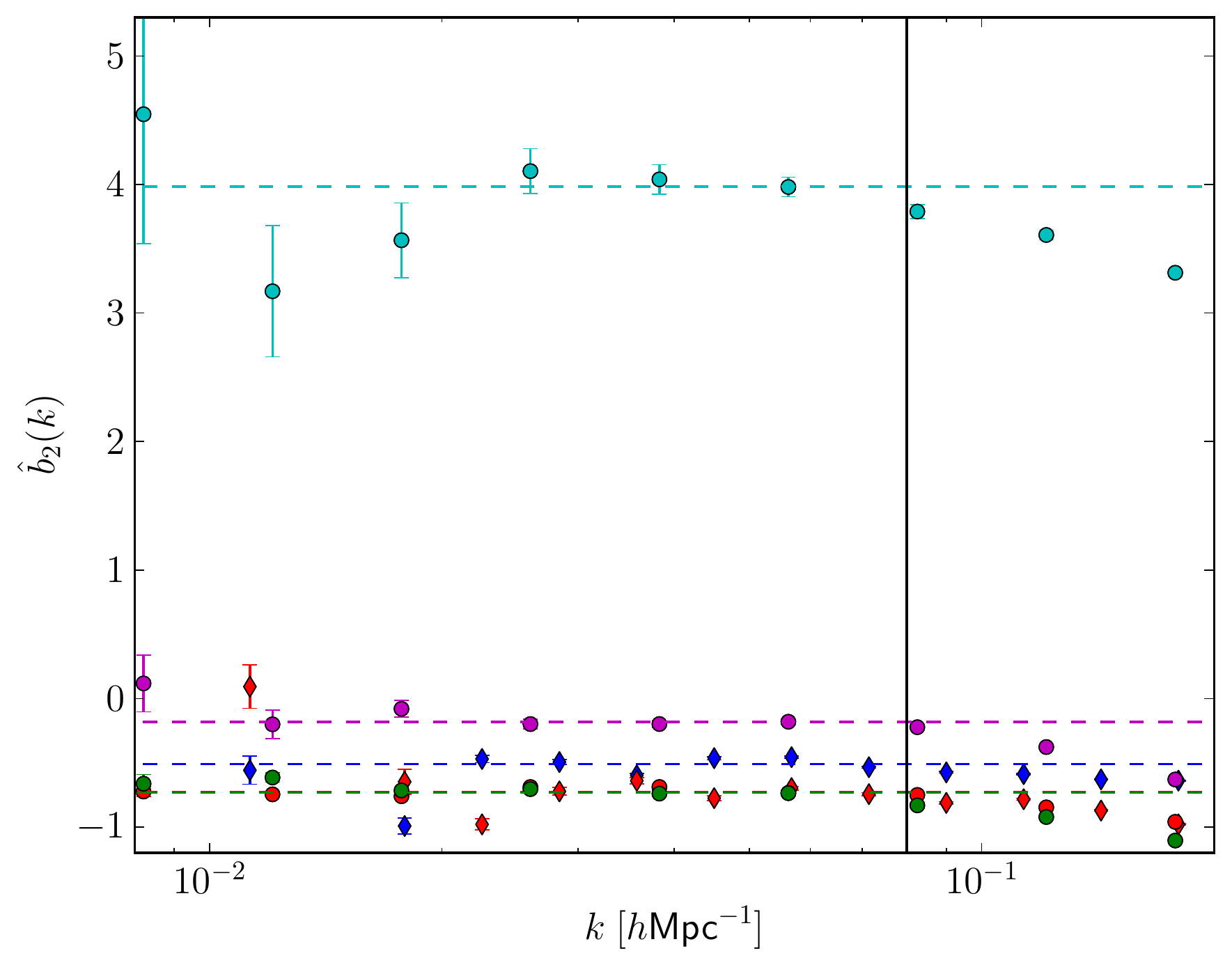}
\includegraphics[width=0.49\textwidth]{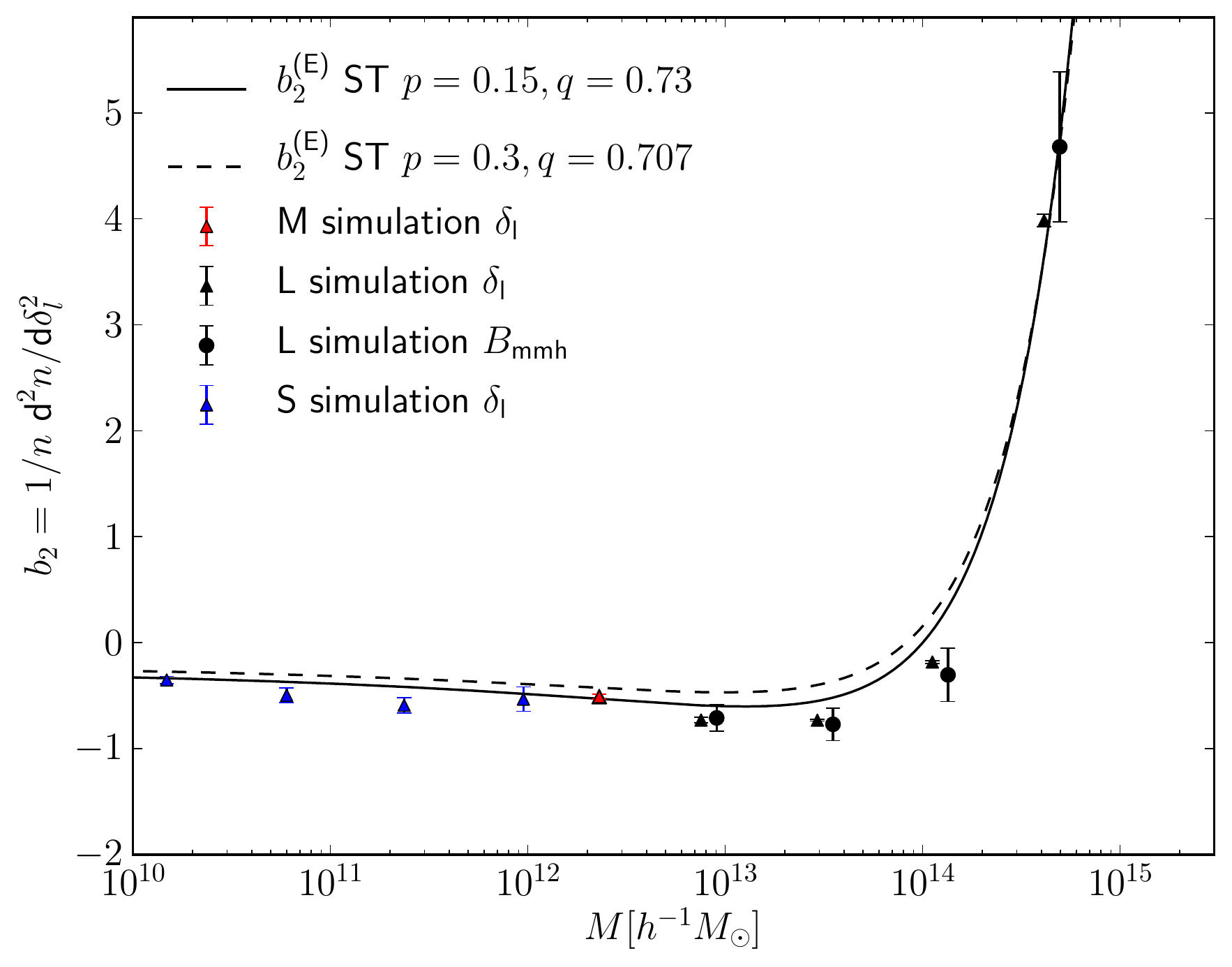}
\caption{Estimation of the second order bias from the clustering in the curved Universe. \emph{Left panel: }Scale dependence of the estimator Eq.~\eqref{eq:b2clustest} and large scale fit. Circles show results from the L simulation and diamonds results from the M simulation. The errorbars are assumed to scale as $1/\sqrt{\bar n N_k P}$, but the overall normalization is left free. The horizontal lines are large scale fits with the weights provided by the error bar and the error of the fit is estimated from the variance of the contributing measurements. \emph{Right panel: }Mass dependence of the large scale fits to $\hat b_2(k)$ (left panel). We are overplotting the second order bias derived from the Sheth-Tormen mass function with two sets of parameters and find reasonable agreement. We also performed a bispectrum analysis following the steps laid out in \cite{Baldauf:2012ev} and find good agreement. Note that the bispectrum has considerably larger error bars.}
\label{fig:b2clust}
\end{figure}

Let us now discuss the halo clustering statistics in the presence of the long wavelength mode. It is useful to define the halo overdensities with respect to the background or global mean,  
\be
\delta_{\text{h},\pm}(\vec x)=\frac{n_{\text{h},\pm}(\vec x)}{\overline n_\text{B}}-1
\label{eq:halodensglobalnorm}
\ee
so that we can compare with perturbative techniques.
The response of the local halo density to a long mode can be derived analogously to Eq.~\eqref{eq:localdenskspace} using~\cite{Baldauf:2012ev,Chan:2011}
\begin{equation}
\begin{split}\label{eq:localhalodenskspace}
\delta_\text{h,G}(\vec k)=&b_1^\text{(E)}\delta^{(1)}(\vec k)+\int \dqc \left\{\frac12 b_2^\text{(E)} +b_1^\text{(E)}F_2(\vec q,\vec k-\vec q)+b_{s^2}S_2(\vec q,\vec k-\vec q)\right\}\delta^{(1)}(\vec q)\delta^{(1)}(\vec k-\vec q)\\
\stackrel{q\ll k}{=}&b_1^\text{(E)} \delta_\text{B}^{(1)}\left[\vec k\left(1-\frac{\delta_\text{l}}{3}\right)\right]+\left(\frac{13}{21}b_1^\text{(E)}+b_2^\text{(E)}\right)\delta_\text{B}^{(1)}(\vec k)\delta_\text{l}
\end{split}
\end{equation}
Comparing to Eq.~\eqref{eq:localdenskspace}, we see that the response of the local halo overdensity to a long mode has an additional dependence on the second order Eulerian bias parameter $b_2^\text{(E)}$. This dependence arises from a coupling of short and long modes in $b_2^\text{(E)}\delta^2$.
Defining the estimated local scale dependent bias as the ratio of the halo-matter and matter power spectra in the curved simulations and using Eq.~\eqref{eq:deltatop1} we get
\be
\hat b_{1,\pm}(k)=\frac{P_{\text{hm},\pm}}{P_{\text{mm},\pm}}=\frac{b_1^\text{(E)}+\frac{47}{21}b_1^\text{(E)} \deltal+b_2^\text{(E)}\deltal }{1+\frac{47}{21}\deltal}\approx b_1^\text{(E)}+b_2^\text{(E)}\deltal, 
\ee
we can therefore estimate the second order bias as
\be\label{eq:b2clustest}
\hat b_2^\text{(E)}=\frac{\hat b_{1,+}(k)-\hat b_{1,-}(k)}{2|\delta_\text{l}|}\;.
\ee
In Fig.~\ref{fig:b2clust} we show the estimator $\hat b_2^\text{(E)}$, which indeed asymptotes to a constant on large scales. The mass dependence of the low-$k$ limit is in reasonable agreement with the second order bias parameter derived from the Sheth-Tormen bias function, as we show in the right panel of Fig.~\ref{fig:b2clust}. We also compare to a simple bispectrum analysis on the matter-matter-halo bispectrum of the fiducial simulation using triangles with wavenumbers up to $k_\text{max}=0.06\ihMpc$ and the method of \cite{Baldauf:2012ev} and find good agreement.\\
If one in turn is interested in the full observable quantity, one has to correct for the fact that the halo overdensity is typically normalized to the local or observed mean number density and thus needs to correct for the change of the denominator in Eq.~\eqref{eq:halodensglobalnorm}~\footnote{When we do the perturbative calculations, quantities are normalized with respect to the background mean. If we instead survey galaxies, we are only sensitive the number density in the observable patch and therefore have to use this as the mean density.}
\be
1+\delta_{\text{h},\pm}(\vec x)=\frac{n_{\text{h},\pm}(\vec x)}{\overline n_{\pm}}=\frac{n_{\text{h},\pm}(\vec x)}{\overline n_\text{f}}(1-b_1^\text{(E)} \delta_\text{l})\; .
\ee
At linear order in $\deltal$, which is the order to which we work in this paper, this correction only affects the $b_1^\text{(E)}\delta$ term in Eq.~\eqref{eq:localhalodenskspace}. While the halo (or galaxy) overdensity can only be 
normalized with respect to local density, the weak lensing observations of dark matter are sensitive to the total mass, hence to global mean density of the Universe.   
\begin{figure}
\centering
\includegraphics[width=0.49\textwidth]{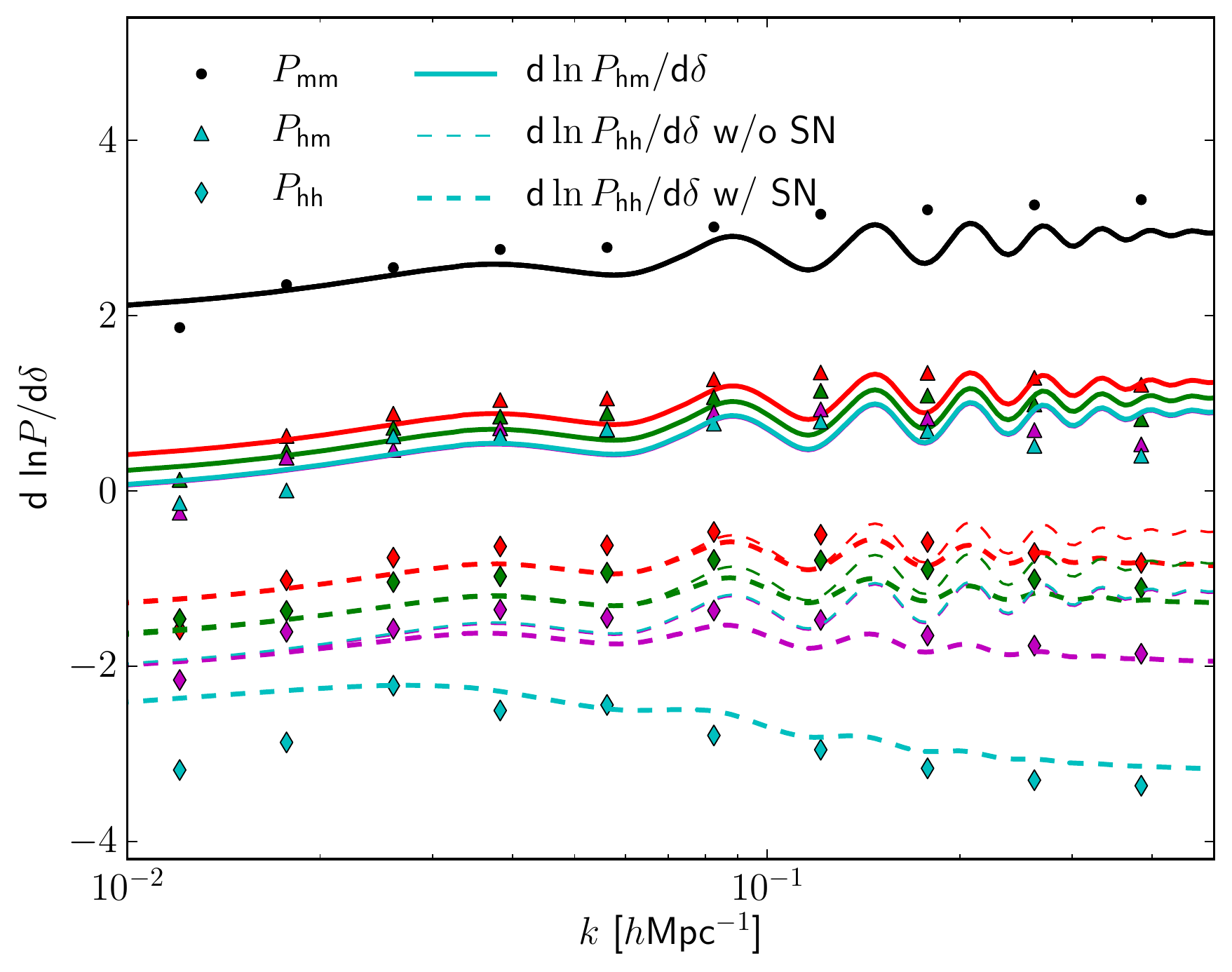}
\caption{Response of matter, halo-matter cross- and halo-halo auto power spectrum to a long wavelength mode.
Black points show the response of the matter power spectrum, colored triangles the response of the halo matter cross power spectrum and colored circles show the response the the halo-halo power spectrum. The mass of haloes is increasing from top to bottom. The slight differences from Fig.~\ref{fig:power} for the mass are related to the coarser binning. The solid and dashed thick lines show the theoretical predictions of Eqs.~\eqref{eq:hmresponse} and \eqref{eq:hhresponse} for halo-matter and halo-halo power spectra. The thin dashed lines show the prediction of \eqref{eq:hhresponse} without the shotnoise corrections.}
\label{fig:response_haloes_full}
\end{figure}
Following Eqs.~\eqref{eq:deltatop1} and \eqref{eq:deltatop2} the response of the halo-matter and halo-halo power spectra to long wavelength fluctuations reads\footnote{Notice that the additional bias terms (beyond the local bias model) present in the complete treatment of bias~\cite{Senatore:2014eva,Mirbabayi:2014zca} do not contribute at this order.}
\be
\begin{split}
\label{eq:hmresponse}
\frac{\derd \ln P_\text{hm}}{\derd \delta_\text{l}}
=&\frac{68}{21}+\frac{b_2^\text{(E)}}{b_1^\text{(E)}}-b_1^\text{(E)}-\frac{1}{3}\frac{\derd \ln k^3 P}{\derd \ln k}
\end{split}
\ee
\be
\begin{split}
\label{eq:hhresponse}
\frac{\derd \ln P_\text{hh}}{\derd \delta_\text{l}}
=&\left(1+\frac{1}{\left[b_1^\text{(E)}\right]^2P \bar n}\right)^{-1}\left(\frac{68}{21}+2\frac{b_2^\text{(E)}}{b_1^\text{(E)}}-2b_1^\text{(E)}-\frac{1}{3}\frac{\derd \ln k^3P}{\derd \ln k}-\frac{1}{b_1^\text{(E)}P \bar n}\right)
\end{split}
\ee
In Fig.~\ref{fig:response_haloes_full} we show the measurements of the response of the halo-matter cross power and halo auto power to the long mode, compare to the above predictions and find good agreement. For the halo matter power spectrum this is a consistency check of the measurement of~(\ref{eq:b2clustest}). Note that the response for halo-halo statistics is negative, mostly due to the normalization to the local mean density.
These ingredients can thus be used to calculate the super sample variance \cite{Li:2014ss} for halo clustering statistics.
The covariance matrix of the real space power spectrum can be written as
\be
\text{Cov}[P_\text{hh}(k_i)P_\text{hh}(k_j)]=P_\text{hh}(k_i)P_\text{hh}(k_j)V^{-1}\left[{4\pi^2 \over k_i^2\Delta k}\delta_{ij}^\text{(K)} + 
\frac{\derd \ln P_\text{hh}(k_i)}{\derd \delta_\text{l}}
\frac{\derd \ln P_\text{hh}(k_j)}{\derd \delta_\text{l}}\sigma^2_V \right],
\ee
where $V$ is the volume of the survey and $\sigma_V$ is the rms variance of density fluctuations on the survey volume.   
The first term describes the disconnected (Gaussian) part of the covariance matrix, while the second is the super-sample variance, 
which is a contribution to the covariance matrix from the modes that do not average to zero within the survey volume. 
The covariance depends on $\sigma_V$, whose typical value is about 0.4\% at $z=0$ for a $1\; h^{-3}\text{Gpc}^3$ volume, and hence the corresponding super-sample variance for this volume,
for a biased tracer with $b_1^\text{(E)}=2$ with $\derd \ln P_\text{hh}/\derd \delta_\text{l} \sim -2$, is about 0.8\%.        
This contribution is only part of the total covariance matrix. In addition there are contributions from the modes within
the survey, as well as from the redshift space distortions. 

%===============================================================================
%===============================================================================
%===============================================================================
%===============================================================================
\section{Bias of non-linear Transformations and Ly-$\alpha$ Forest}\label{sec:nltrans}
We will now consider how non-linear transformations of the density field respond to long wavelength fluctuations. This will in turn allow us to write the two point function of these transformations as a bias times the two point function of the long mode. For this purpose let us start by considering how the $n$-th power of the field responds to a long wavelength fluctuation following \cite{Seljak:2012ja}.\\
%>>>>>>>>>>>>>>>>>>>>>>>>>>>>>>>>>>>>>>>>>>>>>>>>>>>>>>>>>>>>>>>>>>>>>>>>>>>>>>>>>>
\begin{figure}[t]
\begin{center}
\includegraphics[width=0.49\textwidth]{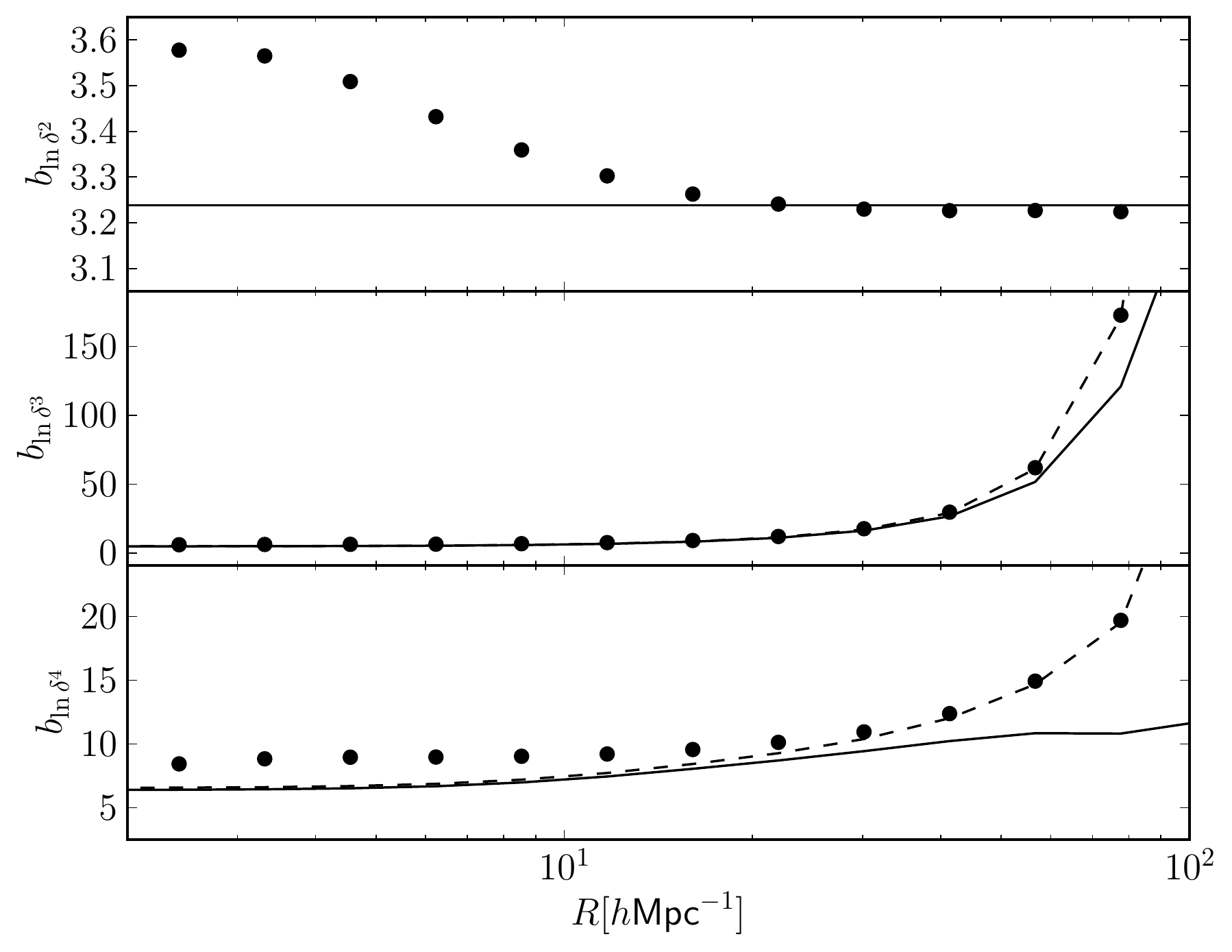}
\caption{Logarithmic bias of the second, third and fourth moment of the density field. The lines are given by the corresponding prediction according to Eq.~\eqref{eq:nltransbias}, solid lines correspond to Eq.~\eqref{eq:nltransbias} and the dashed lines correct this prediction for $\deltal^2$ corrections present in our numerical derivatives.}
\label{fig:mombias}
\end{center}
\end{figure}
%<<<<<<<<<<<<<<<<<<<<<<<<<<<<<<<<<<<<<<<<<<<<<<<<<<<<<<<<<<<<<<<<<<<<<<<<<<<<<<<<<<
As we have seen before in Eq.~\eqref{eq:localdensrescale}, a short wavelength perturbation responds to a long wavelength perturbation as $\delta_\text{s}\to (1+\lambda \delta_\text{l})\delta_\text{s}$, where $\lambda=34/21$. Note that this relation holds only to the leading order in the small non-linearities. Splitting a general perturbation into its short and long components $\delta=\delta_\text{s}+\delta_\text{l}$, we have for its $n$-th power
\be
\delta^n \approx \left(1 + n \lambda \delta_\text{l}\right) \delta_\text{s}^n+n \delta_\text{l} \delta_\text{s}^{n-1} \; .
\label{eq:nlrescal}
\ee
Thus we have for the logarithmic bias 
\be
b_{\ln \delta^n}=\frac{\derd \ln \delta^n}{\derd \delta_\text{l}}\biggr|_{\delta_\text{l}=0}=n \lambda +n  \frac{\la \delta_\text{s}^{n-1}\ra}{\la \delta_\text{s}^n\ra}\; .
\label{eq:nltransbias}
\ee
We can determine the response to the long mode on the lhs from our curvature simulations and the averages over powers of the short mode on the rhs from the fiducial simulation.
Operationally, we interpolate the matter on a grid, smooth the grid with a top hat filter, calculate the non-linear transformation at each grid cell and finally average over all cells.
In Fig.~\ref{fig:mombias} we show the simulation measurements of this bias  and overplot the theoretical prediction given in the above equation. The approach works very well for the variance on large scales as one would have expected based on the power spectrum results in Sec.~\ref{sec:power}. For the skewness and curtosis, higher powers of $\deltal$ in Eq.~\eqref{eq:nlrescal} can play a role in the numerical derivative if $\la \delta_\text{s}^n\ra\approx \deltal^n$, which happens for large smoothing scales. We account for this effect in our plots (dashed lines) but note that this is not a sign of the breakdown of Eq.~\eqref{eq:nltransbias}, but an artifact of calculating the numerical derivative with a finite $\deltal$.\footnote{For definiteness, the quadratic $\deltal^2$ corrections to Eq.~\eqref{eq:nltransbias} arising from cubic terms in Eq~\eqref{eq:nlrescal} show up in
\be
\frac{\derd \ln \delta^n}{\derd \deltal}=\frac{\la\delta^n_+\ra-\la\delta^n_-\ra}{2\deltal \la\delta^n_\text{f}\ra}=n \lambda +n  \frac{\la \delta_\text{s}^{n-1}\ra}{\la \delta_\text{s}^n\ra}+C_{2,n}
\ee
where $C_{2,n}$ is given by
\be
C_{2,3}=\deltal^2\left(\frac{1}{\la \delta_\text{s}^3 \ra} + 3 \frac{\la \delta_\text{s}^2\ra}{\la \delta_\text{s}^3 \ra} \lambda^2 +  \lambda^3\right)\; ,
\ee
for $n=3$ and
\be
C_{2,4}=\deltal^2\left(12 \frac{\la \delta_\text{s}^2 \ra}{\la \delta_\text{s}^4 \ra} \lambda + 12 \frac{\la \delta_\text{s}^3\ra}{\la \delta_\text{s}^4 \ra}  \lambda^2 + 4 \lambda^3\right)
\ee
for $n=4$.
} Once this correction has been implemented, the large scale response is in perfect agreement with Eq.~\eqref{eq:nltransbias}. Going to smaller scales would require the calculation of the response at non-linear level in the short modes.
%>>>>>>>>>>>>>>>>>>>>>>>>>>>>>>>>>>>>>>>>>>>>>>>>>>>>>>>>>>>>>>>>>>>>>>>>>>>>>>>>>>
\begin{figure}[t]
\begin{center}
\includegraphics[width=0.49\textwidth]{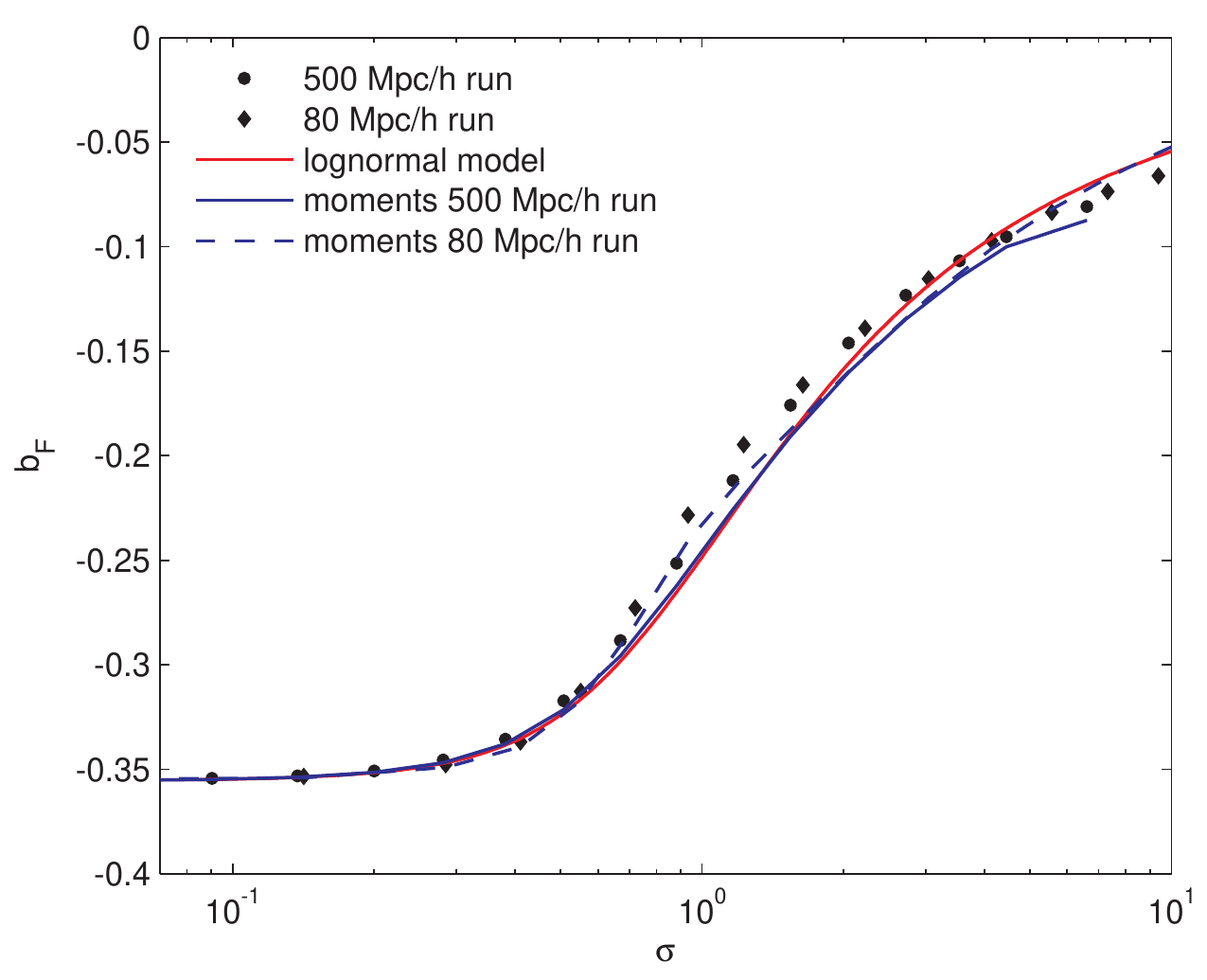}
\caption{
Bias of the flux transformation measured using the curvature method in the S and M simulations. The blue dashed and solid lines show the theoretical prediction with the expectation values in Eq.~\eqref{eq:fluxbias} measured in the S and M simulations, respectively. The red solid line is based on evaluating the expectation values in Eq.~\eqref{eq:fluxbias} in the lognormal model.}
\label{fig:lyalbias}
\end{center}
\end{figure}
%<<<<<<<<<<<<<<<<<<<<<<<<<<<<<<<<<<<<<<<<<<<<<<<<<<<<<<<<<<<<<<<<<<<<<<<<<<<<<<<<<<
Equally good agreement has been found recently by \cite{Cieplak:2015} at higher redshifts.\\
Let us now focus on a specific observable non-linear transformation of the density field, the Ly-$\alpha$ forest. This effect describes the multiple absorption features imprinted into the continuous emission spectrum of background quasars. These absorption features arise if the redshifted photons are in resonance with the Ly-$\alpha$ transition of intervening neutral hydrogen.  The observable, the transmitted flux
as a function of frequency is then related to the emitted flux by the optical depth along the line of sight.
The transmitted flux in the fluctuating Gunn-Peterson approximation \cite{Croft:1998re} is then given by $F=\eh{-\tau}$, where $\tau$ is the
optical depth given by $\tau=A(1+\delta)^\alpha$, where we choose $\alpha=1.6$ and $A=0.3$ for definiteness.
For the response of the mean flux on a background mode one has then
\be
\begin{split}
b_F=\frac{\derd F}{\derd \deltal}=&-A \alpha \lambda   \la  F \delta_\text{s} \left(1+\delta_\text{s}\right)^{\alpha-1}\ra-A \alpha   \la  F  \left(1+\delta_\text{s}\right)^{\alpha-1}\ra\; .%\\
\label{eq:fluxbias}
\end{split}
\ee
As usual, the lhs can be obtained in the curvature approach by measuring $F$ in curved Universe simulation and taking the derivative with respect to $\deltal$. The averages on the right hand side of the above equation can either be measured in the flat simulation or calculated from a theoretical probability density function (PDF). While the initial density field is close to Gaussian, the non-linear clustering will generate a highly non-Gaussian PDF. The observed flux PDF is modeled quite well by a log-normal (LN) distribution \cite{Seljak:2012ja}, which can be generated by transforming a Gaussian density field $\delta_\text{G}$ with dispersion $\sigma_\text{G}$ as
\be
\delta_\text{LN}=\eh{\delta_\text{G}-\frac{\sigma_\text{G}^2}{2}}-1\; ,
\ee
with PDF
\be
p_\text{LN}(\delta_\text{LN})\derd \delta_\text{LN}=\frac{1}{1+\delta_\text{NL}}\frac{1}{\sqrt{2\pi \sigma^2}}\eh{-\frac12\frac{\bigl(\ln (1+\delta_\text{LN})+\sigma_\text{G}^2/2\bigr)^2}{\sigma_\text{G}^2}}
\ee
and support $[-1,+\infty]$. The mean is zero, the variance is $\eh{\sigma_\text{G}^2}-1$.
In Fig.~\ref{fig:lyalbias} we show the flux bias inferred from the curvature simulations as a function of the variance of the underlying field.
We see that the feedback of the flux transformation on the presence of a background mode is very well described by Eq.~\eqref{eq:fluxbias} with the expectation values on the right hand side measured in the same box but $\deltal=0$. The expectation values can also be estimated from the lognormal model, yielding equally good agreement. 
A similar approach has been used by \cite{McDonald:2003fx} and \cite{Cieplak:2015} in smaller scale simulations, reporting good results in in the absence of thermal effects, but the agreement is considerably worse when those are included \cite{Wang:2015op}. 

%===============================================================================
%===============================================================================

%===============================================================================
%===============================================================================
\section{Summary and Discussion}\label{sec:concl}
This paper implements the curvature simulations proposed in \cite{Baldauf:2011ew}. We show that the bias inferred from the numerical derivative of the halo number in the curvature simulations with respect to the  long wavelength density perturbation agrees quite well with the bias inferred from the power spectrum. The big advantage of our new method is that it is limited by Poisson errors only (which are reduced by the fact that we can use large primordial fluctuations) and not by large scale cosmic variance. For a fixed error bar, this allows us to measure the bias in comparably smaller volume $N$-body simulations. An additional advantage of our method is, that it automatically measures the infinite distance limit of the bias parameters and is thus safe from loop and higher derivative contaminations present in the standard power spectrum method. Our results agree reasonably well with the ST mass function for masses exceeding $5 \times 10^{12} \hMs$, but at the low mass end the ST prediction seems to considerably underpredict the bias function.

Furthermore, we studied the response of the halo mass function with respect to changes in the fluctuation amplitude $\sigma_8$. This is of particular importance for studies of primordial non-Gaussianity, where the coupling between long- and short wavelength fluctuations leads to a local modulation of the fluctuation amplitude by the long wavelength gravitational potential. We show that, for the FoF haloes employed here, there is clear evidence for deviations from the commonly employed relation between the density and potential bias that leads to a $18\%$ change in the inferred non-Gaussianity amplitude $f_\text{NL}$.

Besides the response of the local number density of haloes, the curvature method also allows us to consider the response of the power spectra to long modes. For the matter power spectra this has been done in \cite{Li:2014ss,Wagner:2015th,Wagner:2015se}, but here we extend this study to the halo power spectra (and their cross-correlation with matter), 
for which additional terms proportional to the first and second order Eulerian bias arise. This provides us with a way to estimate the second order bias, for which we find good agreement with the bispectrum and with the ST mass function. Furthermore, this allows us to verify the newly derived expressions for the super-sample variance of halo two point statistics. For example, we
find that the super-sample variance response is negative for the high bias halos, and is of order $\derd \ln P/\derd \deltal=-2$ for $b_1^\text{(E)}=2$. This method could be extended to the bispectrum in the curved simulations, which probes the squeezed trispectrum. This might provide an efficient way to estimate non-local cubic bias parameters without having to resort to loop corrections \cite{Saito:2014un}. Generally, measuring $N$-point functions in the curved Universe allows one to probe the squeezed $(N+1)$-point function.

Finally, we consider the response of non-linear transformations and Ly-$\alpha$ bias. For large smoothing scales the response of the first three moments of the field to a long mode is in good agreement with the peak-background split derivation of \cite{Seljak:2012ja}. Similarly to \cite{Cieplak:2015}, we find excellent agreement for the density bias of the Ly-$\alpha$ forest flux.

%===============================================================================
%				ACKNOWEDGEMENTS
%===============================================================================
\acknowledgments
We would like to thank A.~Font-Ribera, Y. Li, and E. Schaan for discussions, and V.~Desjacques for comments and initial help with the simulations.
T.B. would like to thank the Berkeley Center for Cosmological Physics and the Lawrence Berkeley Laboratory for the kind hospitality. The simulations were performed on the ZBOX3 supercomputer of the Institute for Theoretical Physics at the University of Zurich. T.B. gratefully acknowledges support from the Institute for Advanced Study through a Corning Glass Works Foundation grant. U.S. acknowledges support from NASA grant NNX15AL17G. L.S. is supported by DOE Early Career Award DE-FG02-12ER41854 and by NSF grant PHY-1068380. M.Z. is supported in part by the NSF grants PHY-1213563,  PHY-1521097 and AST-1409709.

\bibliographystyle{JHEP}

\bibliography{simulations}

\providecommand{\href}[2]{#2}\begingroup\raggedright\begin{thebibliography}{10}

\bibitem{Dalal:2008hh}
N.~{Dalal}, O.~{Dor{\'e}}, D.~{Huterer}, and A.~{Shirokov}, {\it {Imprints of
  primordial non-Gaussianities on large-scale structure: Scale-dependent bias
  and abundance of virialized objects}},  {\em \prd} {\bf 77} (June, 2008)
  123514, [\href{http://xxx.lanl.gov/abs/0710.4560}{{\tt arXiv:0710.4560}}].

\bibitem{Baldauf:2011ew}
T.~{Baldauf}, U.~{Seljak}, L.~{Senatore}, and M.~{Zaldarriaga}, {\it {Galaxy
  bias and non-linear structure formation in general relativity}},  {\em \jcap}
  {\bf 10} (Oct., 2011) 31, [\href{http://xxx.lanl.gov/abs/1106.5507}{{\tt
  arXiv:1106.5507}}].

\bibitem{Wagner:2015se}
C.~{Wagner}, F.~{Schmidt}, C.-T. {Chiang}, and E.~{Komatsu}, {\it {Separate
  universe simulations}},  {\em \mnras} {\bf 448} (Mar., 2015) L11--L15,
  [\href{http://xxx.lanl.gov/abs/1409.6294}{{\tt arXiv:1409.6294}}].

\bibitem{Wagner:2015th}
C.~{Wagner}, F.~{Schmidt}, C.-T. {Chiang}, and E.~{Komatsu}, {\it {The
  angle-averaged squeezed limit of nonlinear matter N-point functions}},  {\em
  \jcap} {\bf 8} (Aug., 2015) 42,
  [\href{http://xxx.lanl.gov/abs/1503.0348}{{\tt arXiv:1503.0348}}].

\bibitem{McDonald:2003fx}
P.~{McDonald}, {\it {Toward a Measurement of the Cosmological Geometry at z
  \~{} 2: Predicting Ly{$\alpha$} Forest Correlation in Three Dimensions and
  the Potential of Future Data Sets}},  {\em \apj} {\bf 585} (Mar., 2003)
  34--51, [\href{http://xxx.lanl.gov/abs/astro-ph/}{{\tt astro-ph/}}].

\bibitem{Seljak:2012ja}
U.~{Seljak}, {\it {Bias, redshift space distortions and primordial
  nongaussianity of nonlinear transformations: application to Lyman alpha
  forest}},  {\em ArXiv e-prints} (Jan., 2012)
  [\href{http://xxx.lanl.gov/abs/1201.0594}{{\tt arXiv:1201.0594}}].

\bibitem{Sirko:2005in}
E.~{Sirko}, {\it {Initial Conditions to Cosmological N-Body Simulations, or,
  How to Run an Ensemble of Simulations}},  {\em \apj} {\bf 634} (Nov., 2005)
  728--743, [\href{http://xxx.lanl.gov/abs/astro-ph/0503106}{{\tt
  astro-ph/0503106}}].

\bibitem{Li:2014ss}
Y.~{Li}, W.~{Hu}, and M.~{Takada}, {\it {Super-sample covariance in
  simulations}},  {\em \prd} {\bf 89} (Apr., 2014) 083519,
  [\href{http://xxx.lanl.gov/abs/1401.0385}{{\tt arXiv:1401.0385}}].

\bibitem{Seljak:1996ap}
U.~{Seljak} and M.~{Zaldarriaga}, {\it {A Line-of-Sight Integration Approach to
  Cosmic Microwave Background Anisotropies}},  {\em \apj} {\bf 469} (Oct.,
  1996) 437, [\href{http://xxx.lanl.gov/abs/astro-ph/}{{\tt astro-ph/}}].

\bibitem{Springel:2005mi}
V.~Springel, {\it {The cosmological simulation code GADGET-2}},  {\em Mon. Not.
  Roy. Astron. Soc.} {\bf 364} (2005) 1105--1134,
  [\href{http://xxx.lanl.gov/abs/astro-ph/0505010}{{\tt astro-ph/0505010}}].

\bibitem{Komatsu:2008hk}
{\bf WMAP} Collaboration, E.~Komatsu {\em et.~al.}, {\it {Five-Year Wilkinson
  Microwave Anisotropy Probe (WMAP) Observations:Cosmological Interpretation}},
   {\em Astrophys. J. Suppl.} {\bf 180} (2009) 330--376,
  [\href{http://xxx.lanl.gov/abs/0803.0547}{{\tt arXiv:0803.0547}}].

\bibitem{Creminelli:2013cga}
P.~{Creminelli}, A.~{Perko}, L.~{Senatore}, M.~{Simonovi{\'c}}, and
  G.~{Trevisan}, {\it {The physical squeezed limit: consistency relations at
  order q$^{2}$}},  {\em \jcap} {\bf 11} (Nov., 2013) 15,
  [\href{http://xxx.lanl.gov/abs/1307.0503}{{\tt arXiv:1307.0503}}].

\bibitem{Bernardeau:2002}
F.~{Bernardeau}, S.~{Colombi}, E.~{Gazta{\~n}aga}, and R.~{Scoccimarro}, {\it
  {Large-scale structure of the Universe and cosmological perturbation
  theory}},  {\em \physrep} {\bf 367} (Sept., 2002) 1--248,
  [\href{http://xxx.lanl.gov/abs/astro-ph/0112551}{{\tt astro-ph/0112551}}].

\bibitem{Takahashi:2008yk}
R.~Takahashi, {\it {Third Order Density Perturbation and One-loop Power
  Spectrum in a Dark Energy Dominated Universe}},  {\em Prog.Theor.Phys.} {\bf
  120} (2008) 549--559, [\href{http://xxx.lanl.gov/abs/0806.1437}{{\tt
  arXiv:0806.1437}}].

\bibitem{Sherwin:2012ar}
B.~D. {Sherwin} and M.~{Zaldarriaga}, {\it {The Shift of the Baryon Acoustic
  Oscillation Scale: A Simple Physical Picture}},  {\em ArXiv e-prints} (Feb.,
  2012) [\href{http://xxx.lanl.gov/abs/1202.3998}{{\tt arXiv:1202.3998}}].

\bibitem{Senatore:2015ir}
L.~{Senatore} and M.~{Zaldarriaga}, {\it {The IR-resummed Effective Field
  Theory of Large Scale Structures}},  {\em \jcap} {\bf 2} (Feb., 2015) 13,
  [\href{http://xxx.lanl.gov/abs/1404.5954}{{\tt arXiv:1404.5954}}].

\bibitem{Baldauf:2015eq}
T.~{Baldauf}, M.~{Mirbabayi}, M.~{Simonovi{\'c}}, and M.~{Zaldarriaga}, {\it
  {Equivalence principle and the baryon acoustic peak}},  {\em \prd} {\bf 92}
  (Aug., 2015) 043514, [\href{http://xxx.lanl.gov/abs/1504.0436}{{\tt
  arXiv:1504.0436}}].

\bibitem{McDonald:2009dh}
P.~{McDonald} and A.~{Roy}, {\it {Clustering of dark matter tracers:
  generalizing bias for the coming era of precision LSS}},  {\em \jcap} {\bf 8}
  (Aug., 2009) 20, [\href{http://xxx.lanl.gov/abs/0902.0991}{{\tt
  arXiv:0902.0991}}].

\bibitem{Assassi:2014re}
V.~{Assassi}, D.~{Baumann}, D.~{Green}, and M.~{Zaldarriaga}, {\it
  {Renormalized halo bias}},  {\em \jcap} {\bf 8} (Aug., 2014) 56,
  [\href{http://xxx.lanl.gov/abs/1402.5916}{{\tt arXiv:1402.5916}}].

\bibitem{Senatore:2014eva}
L.~{Senatore}, {\it {Bias in the Effective Field Theory of Large Scale
  Structures}},  {\em ArXiv e-prints} (June, 2014)
  [\href{http://xxx.lanl.gov/abs/1406.7843}{{\tt arXiv:1406.7843}}].

\bibitem{Mirbabayi:2014zca}
M.~{Mirbabayi}, F.~{Schmidt}, and M.~{Zaldarriaga}, {\it {Biased tracers and
  time evolution}},  {\em \jcap} {\bf 7} (July, 2015) 30,
  [\href{http://xxx.lanl.gov/abs/1412.5169}{{\tt arXiv:1412.5169}}].

\bibitem{Lewandowski:2015an}
M.~{Lewandowski}, A.~{Perko}, and L.~{Senatore}, {\it {Analytic prediction of
  baryonic effects from the EFT of large scale structures}},  {\em \jcap} {\bf
  5} (May, 2015) 19, [\href{http://xxx.lanl.gov/abs/1412.5049}{{\tt
  arXiv:1412.5049}}].

\bibitem{Angulo:2015eqa}
R.~{Angulo}, M.~{Fasiello}, L.~{Senatore}, and Z.~{Vlah}, {\it {On the
  statistics of biased tracers in the Effective Field Theory of Large Scale
  Structures}},  {\em \jcap} {\bf 9} (Sept., 2015) 29,
  [\href{http://xxx.lanl.gov/abs/1503.0882}{{\tt arXiv:1503.0882}}].

\bibitem{Assassi:2015ef}
V.~{Assassi}, D.~{Baumann}, E.~{Pajer}, Y.~{Welling}, and D.~{van der Woude},
  {\it {Effective Theory of Large-Scale Structure with Primordial
  Non-Gaussianity}},  {\em ArXiv e-prints} (May, 2015)
  [\href{http://xxx.lanl.gov/abs/1505.0666}{{\tt arXiv:1505.0666}}].

\bibitem{Assassi:2015ga}
V.~{Assassi}, D.~{Baumann}, and F.~{Schmidt}, {\it {Galaxy Bias and Primordial
  Non-Gaussianity}},  {\em ArXiv e-prints} (Oct., 2015)
  [\href{http://xxx.lanl.gov/abs/1510.0372}{{\tt arXiv:1510.0372}}].

\bibitem{Mo:1996an}
H.~J. {Mo} and S.~D.~M. {White}, {\it {An analytic model for the spatial
  clustering of dark matter haloes}},  {\em \mnras} {\bf 282} (Sept., 1996)
  347--361, [\href{http://xxx.lanl.gov/abs/astro-ph/}{{\tt astro-ph/}}].

\bibitem{Press:1974fo}
W.~H. {Press} and P.~{Schechter}, {\it {Formation of galaxies and clusters of
  galaxies by selfsimilar gravitational condensation}},  {\em Astrophys. J.}
  {\bf 187} (1974) 425--438.

\bibitem{Sheth:1999mn}
R.~K. Sheth and G.~Tormen, {\it {Large scale bias and the peak background
  split}},  {\em Mon.Not.Roy.Astron.Soc.} {\bf 308} (1999) 119,
  [\href{http://xxx.lanl.gov/abs/astro-ph/9901122}{{\tt astro-ph/9901122}}].

\bibitem{Tinker:2008to}
J.~{Tinker}, A.~V. {Kravtsov}, A.~{Klypin}, K.~{Abazajian}, M.~{Warren},
  G.~{Yepes}, S.~{Gottl{\"o}ber}, and D.~E. {Holz}, {\it {Toward a Halo Mass
  Function for Precision Cosmology: The Limits of Universality}},  {\em \apj}
  {\bf 688} (Dec., 2008) 709--728,
  [\href{http://xxx.lanl.gov/abs/0803.2706}{{\tt arXiv:0803.2706}}].

\bibitem{Paranjape:2013ex}
A.~{Paranjape}, R.~K. {Sheth}, and V.~{Desjacques}, {\it {Excursion set peaks:
  a self-consistent model of dark halo abundances and clustering}},  {\em
  \mnras} {\bf 431} (May, 2013) 1503--1512,
  [\href{http://xxx.lanl.gov/abs/1210.1483}{{\tt arXiv:1210.1483}}].

\bibitem{Seljak:2004mn}
U.~{Seljak} and M.~S. {Warren}, {\it {Large-scale bias and stochasticity of
  haloes and dark matter}},  {\em \mnras} {\bf 355} (Nov., 2004) 129--136,
  [\href{http://xxx.lanl.gov/abs/astro-ph/}{{\tt astro-ph/}}].

\bibitem{Grossi:2009mn}
M.~{Grossi}, L.~{Verde}, C.~{Carbone}, K.~{Dolag}, E.~{Branchini},
  F.~{Iannuzzi}, S.~{Matarrese}, and L.~{Moscardini}, {\it {Large-scale
  non-Gaussian mass function and halo bias: tests on N-body simulations}},
  {\em \mnras} {\bf 398} (Sept., 2009) 321--332,
  [\href{http://xxx.lanl.gov/abs/0902.2013}{{\tt arXiv:0902.2013}}].

\bibitem{Desjacques:2010pr}
V.~{Desjacques} and U.~{Seljak}, {\it {Primordial Non-Gaussianity in the
  Large-Scale Structure of the Universe}},  {\em Advances in Astronomy} {\bf
  2010} (2010) [\href{http://xxx.lanl.gov/abs/1006.4763}{{\tt
  arXiv:1006.4763}}].

\bibitem{Giannantonio:2010}
T.~{Giannantonio} and C.~{Porciani}, {\it {Structure formation from
  non-Gaussian initial conditions: Multivariate biasing, statistics, and
  comparison with N-body simulations}},  {\em \prd} {\bf 81} (Mar., 2010)
  063530, [\href{http://xxx.lanl.gov/abs/0911.0017}{{\tt arXiv:0911.0017}}].

\bibitem{Hamaus:2011op}
N.~{Hamaus}, U.~{Seljak}, and V.~{Desjacques}, {\it {Optimal constraints on
  local primordial non-Gaussianity from the two-point statistics of large-scale
  structure}},  {\em \prd} {\bf 84} (Oct., 2011) 083509,
  [\href{http://xxx.lanl.gov/abs/1104.2321}{{\tt arXiv:1104.2321}}].

\bibitem{Saito:2014un}
S.~{Saito}, T.~{Baldauf}, Z.~{Vlah}, U.~{Seljak}, T.~{Okumura}, and
  P.~{McDonald}, {\it {Understanding higher-order nonlocal halo bias at large
  scales by combining the power spectrum with the bispectrum}},  {\em ArXiv
  e-prints} (May, 2014) [\href{http://xxx.lanl.gov/abs/1405.1447}{{\tt
  arXiv:1405.1447}}].

\bibitem{Manera:2010la}
M.~{Manera}, R.~K. {Sheth}, and R.~{Scoccimarro}, {\it {Large-scale bias and
  the inaccuracy of the peak-background split}},  {\em \mnras} {\bf 402} (Feb.,
  2010) 589--602, [\href{http://xxx.lanl.gov/abs/0906.1314}{{\tt
  arXiv:0906.1314}}].

\bibitem{Fry:1992vr}
J.~N. Fry and E.~Gaztanaga, {\it {Biasing and hierarchical statistics in large
  scale structure}},  {\em Astrophys. J.} {\bf 413} (1993) 447--452,
  [\href{http://xxx.lanl.gov/abs/astro-ph/9302009}{{\tt astro-ph/9302009}}].

\bibitem{McDonald:2006cl}
P.~{McDonald}, {\it {Clustering of dark matter tracers: Renormalizing the bias
  parameters}},  {\em \prd} {\bf 74} (Nov., 2006) 103512,
  [\href{http://xxx.lanl.gov/abs/astro-ph/}{{\tt astro-ph/}}].

\bibitem{Schmidt:2013pe}
F.~{Schmidt}, D.~{Jeong}, and V.~{Desjacques}, {\it {Peak-background split,
  renormalization, and galaxy clustering}},  {\em \prd} {\bf 88} (July, 2013)
  023515, [\href{http://xxx.lanl.gov/abs/1212.0868}{{\tt arXiv:1212.0868}}].

\bibitem{Baldauf:2012ev}
T.~{Baldauf}, U.~{Seljak}, V.~{Desjacques}, and P.~{McDonald}, {\it {Evidence
  for quadratic tidal tensor bias from the halo bispectrum}},  {\em \prd} {\bf
  86} (Oct., 2012) 083540, [\href{http://xxx.lanl.gov/abs/1201.4827}{{\tt
  arXiv:1201.4827}}].

\bibitem{Chan:2011}
K.~C. {Chan}, R.~{Scoccimarro}, and R.~K. {Sheth}, {\it {Gravity and
  large-scale nonlocal bias}},  {\em \prd} {\bf 85} (Apr., 2012) 083509,
  [\href{http://xxx.lanl.gov/abs/1201.3614}{{\tt arXiv:1201.3614}}].

\bibitem{Cieplak:2015}
A.~M. {Cieplak} and A.~{Slosar}, {\it {Towards physics responsible for
  large-scale Lyman-\$$\backslash$alpha\$ forest bias parameters}},  {\em ArXiv
  e-prints} (Sept., 2015) [\href{http://xxx.lanl.gov/abs/1509.0787}{{\tt
  arXiv:1509.0787}}].

\bibitem{Croft:1998re}
R.~A.~C. {Croft}, D.~H. {Weinberg}, N.~{Katz}, and L.~{Hernquist}, {\it
  {Recovery of the Power Spectrum of Mass Fluctuations from Observations of the
  Ly{$\alpha$} Forest}},  {\em \apj} {\bf 495} (Mar., 1998) 44--62,
  [\href{http://xxx.lanl.gov/abs/astro-ph/9708018}{{\tt astro-ph/9708018}}].

\bibitem{Wang:2015op}
X.~{Wang}, A.~{Font-Ribera}, and U.~{Seljak}, {\it {Optimizing BAO measurements
  with non-linear transformations of the Lyman-{$\alpha$} forest}},  {\em
  \jcap} {\bf 4} (Apr., 2015) 9, [\href{http://xxx.lanl.gov/abs/1412.4727}{{\tt
  arXiv:1412.4727}}].

\end{thebibliography}\endgroup

\end{document}